\documentclass[12pt,onecolumn]{aa}

\usepackage{lscape}
\usepackage{graphicx,amssymb,amsmath,times}
\usepackage{longtable}
\usepackage{bm} 
\usepackage{url}
\usepackage[T1]{fontenc}
\usepackage{natbib}
\bibpunct{(}{)}{;}{a}{}{,} % to follow the A&A style
\usepackage[normalem]{ulem}

\def\simlt{\lower.5ex\hbox{\ltsima}}
\def\simgt{\lower.5ex\hbox{\gtsima}}

\def\gtsim{\;\lower.6ex\hbox{$\sim$}\kern-6.7pt\raise.4ex\hbox{$>$}\;}
\def\ltsim{\;\lower.6ex\hbox{$\sim$}\kern-6.9pt\raise.4ex\hbox{$<$}\;}

\def\Deg{${}^\circ$\llap{.}}

\def\deg{${}^\circ$}

\def\jmk{\hbox{\it J--K\/}}

\titlerunning{Type II Cepheids in the Galactic Center based on Near-Infrared data}
\authorrunning{Braga et al.}
\usepackage{color}

\begin{document}

\title{New type II Cepheids from VVV data towards the Galactic 
center\thanks{Table 2 is only available in electronic form
at the CDS via anonymous ftp to \protect\url{cdsarc.u-strasbg.fr} (130.79.128.5)
or via \protect\url{http://cdsweb.u-strasbg.fr/cgi-bin/qcat?J/A+A/xxx}}}

% ---Gruppo principale---
   \author{V.~F.~Braga\inst{1,2} 
   \and R.~Contreras Ramos\inst{1,3}
   \and D.~Minniti\inst{1,2,4} 
   \and C.~E.~Ferreira Lopes\inst{5}
   \and M.~Catelan\inst{1,3}\fnmsep\thanks{on sabbatical leave at European Southern Observatory, Alonso de C{\'o}rdova 3107, Casilla 19001, Santiago, Chile}
   \and J.~H.~Minniti\inst{1,3,6}
   \and F.~Nikzat\inst{1,3}
   \and M.~Zoccali\inst{1,3}}

%    \footnote{On sabbatical leave at European Southern Observatory, Av. Alonso de C\'ordova 3107, Vitacura, Santiago Chile}
   
\institute{Instituto Milenio de Astrof{\'i}sica, Santiago, Chile\\
\and
Departamento de F{\'i}sica, Facultad de Ciencias Exactas, Universidad Andr{\'e}s Bello, Fern{\'a}ndez Concha 700, Las Condes, Santiago, Chile\\
\and
Pontificia Universidad Cat{\'o}lica de Chile, Instituto de Astrof{\'i}sica, Av. Vicu\~na Mackenna 4860, 7820436, Macul, Santiago, Chile\\
\and
Vatican Observatory, V00120 Vatican City State, Italy\\
\and
National Institute For Space Research (INPE/MCTI), Av. dos Astronautas, 1758 -- S{\~a}o Jos{\'e} dos Campos -- SP, 12227-010, Brazil\\
\and
European Southern Observatory, Alonso de C{\'o}rdova 3107, Casilla 19001, Santiago, Chile}

\date{\centering Submitted \today\ / Received / Accepted }

\abstract{The Galactic center (GC) is the densest region of the Milky Way. 
Variability surveys towards the GC potentially provide the largest number
of variable stars per square degree within the Galaxy. However, high stellar 
density is also a drawback due to blending. Moreover, the GC is affected
by extreme reddening, therefore near infrared observations are needed.}
{We plan to detect new variable stars towards the GC, focusing on
type II Cepheids (T2Cs) which have the advantage of being brighter 
than RR Lyrae stars.}
{We perform parallel Lomb-Scargle and Generalized Lomb-Scargle 
periodogram analysis of the $K_s$-band time series of the 
VISTA variables in the V{\'i}a L{\'a}ctea 
survey, to detect periodicities. We employ statistical parameters
to clean our sample. We take account of periods, light amplitudes, 
distances, and proper motions to provide a classification of the 
candidate variables.}
{We detected 1,019 periodic variable stars, 
of which 164 are T2Cs, 210
are Miras and 3  are classical Cepheids. We also found the first anomalous 
Cepheid in this region. We compare their photometric properties 
with overlapping catalogs and discuss their properties on the 
color-magnitude and Bailey diagrams.}
{We present the most extensive catalog of T2Cs in the GC region to date.
Offsets in E($J-K_s$) and in the reddening law cause very large
($\sim$1-2 kpc) uncertainties on distances in this region.
We provide a catalog which will be the starting point for future
spectroscopic surveys in the innermost regions of the Galaxy.}

\maketitle

\section{Introduction}

To investigate the central regions of the 
Galaxy is a top priority for studies of Galaxy evolution, since 
the Bulge contains $\sim$40\% of the stellar mass of the Galaxy 
\citep{valenti2016,mcmillan2017}.
However, the Galactic center (GC) is one of the most complicated regions to study 
in the Milky Way because of high extinction and high crowding.
At low galactic latitudes, the optical surveys such as {\it Gaia}
\citep{gaia_alldr}, OGLE \citep{udalski92}, and ASAS \citep{pojmanski1997} 
suffer from extreme reddening and need to be 
complemented with near infrared (NIR) surveys like the 
Two Micron All-Sky Survey \citep[2MASS,][]{skrutskie2006} and 
the VISTA Variables in the V{\'i}a L{\'a}ctea 
\citep[VVV,][]{minniti2010,saito2012} to detect sources in the regions of 
the Galactic plane where visual absorption can be as 
high as $A_V \sim$30 mag, or higher. 

Over the last 10 years the synergy between OGLE and VVV has provided
unprecedented results concerning variable stars in the Bulge thanks
to the exploitation of their advantages and compensation for their disadvantages. 
In fact, OGLE scanned vast regions of the sky, generating
time series with thousands of epochs, and providing the largest 
known catalogs of variable stars in the Bulge and Magellanic Clouds. 
However, the surface density of both periodic variables (RR Lyraes 
[RRLs], Cepheids, Miras...) and transients (microlensing events) drops dramatically 
at low galactic latitudes ($|b|<$1.5\deg). On the other hand, VVV 
could not achieve the same sky area coverage and number of epochs 
due to the intrinsically more telescope time-demanding observation
strategies in the NIR bands. Nonetheless, its deeper
images allowed the detection of Cepheids, RRLs, and microlensing events 
in regions where none had been found before 
\citep[][ henceforth, CR18]{dekany2013,gran2016,
minniti2016,minniti2017b,navarro2017,
majaess2018,navarro2018,contreras2018},
also leading to the discovery of new globular clusters
\citep{minniti2017c,minniti2017d,bica2018}. 
Concerning pulsating variable stars in the instability strip (IS), which are
valuable standard candles since they obey period--luminosity (PL)
relations, the synergy between OGLE and VVV was
discussed by \citet{pietrukowicz2012} 
and exploited by \citet{bhardwaj17c} and \citep{braga2018b} for 
type II Cepheids (T2Cs). However, since these studies were based on the OGLE lists of 
variables,  the inner regions of the Bulge could not be inspected.

In this framework, a census of T2Cs at low latitudes is still missing. These 
variables are not as popular as classical Cepheids (CCs) or RRLs as standard candles.
In fact, they are $\sim$2-3 mag fainter than CCs at the same period --  meaning that they cannot be 
used to estimate distances of stellar systems outside the Local Group -- but they
are also much less numerous than RRLs \citep[less than 1,000 were found in 
the Bulge by OGLE, while the same survey detected 
more than 38,000 RRLs][]{soszynski14,soszynski2017}. On the other hand, 
T2Cs are brighter than RRLs. Moreover, the majority of the 
T2Cs are old ($>$10 Gyr) stars, although it was recently suggested 
that the W Vir (WV) subclass might partly be associated to the 
intermediate-age population \citep{iwanek2018}. This means that 
they can also be found in stellar systems with no recent star 
formation events, which is a requirement to find CCs as they are 
purely young \citep[$<$400 Myr,][]{bono05,anderson2016} 
stars. For a recent review of the 
properties of T2Cs, see the 
monography by \citet{catelan15}.

The aim of this paper is to find new variables in the central-most
tile of the VVV survey, focusing on the detection of T2Cs.
The paper is organized as follows: in Section 2, we present the data. 
In Section 3, we describe the light-curve analysis, including the variable 
star search, period determinations, and light-curve fittings.
In Section 4, we discuss the classification of periodic variable stars 
using all tools available. We present the matches with other 
existing catalogs of GC variable stars in Section 5.
In Section 6, we discuss the position, color-magnitude diagrams
(CMDs) and Bailey (amplitude vs. 
period) diagrams of the final list of variables, 
and the obtained distances to T2Cs.
Finally, the conclusions are presented in Section 7.

\section{Data}\label{sect_data}

{\it Photometry} --- We used proprietary PSF photometric reduction \citep{contreras2017} 
of the VVV data \citep{minniti2010} in tile b333 (--1\Deg23<$l$<0\Deg23; 
--0\Deg46<$b$<0\Deg65), which covers 1.501 deg$^2$ on the 
GC. The camera 
VIRCAM has 16 chips which cover one full tile with 
six pointings, which are referred to as ``pawprints''. Overall, the observation of 
each tile consists of 96 fields of view. We refer to each of these fields of view as a ``pawprint-chip 
combination'' (PCC) . 
The ID of each PCC is a three-digit number, where the 
first digit is the pawprint number, and the second 
and third are the chip number. We note that since there are 
overlaps between the different PCCs, one tile is fully 
covered by 48 PCCs. 
The advantages of PSF-fitting photometry with respect
to the public aperture photometry available at VSA 
(\protect\url{http://horus.roe.ac.uk/vsa/}) are manifold in crowded regions
like b333, which includes the GC; it allows for
blended sources to be resolved more easily, performs a better sky subtraction, 
and provides a deeper limit magnitude (at least $\sim$2 mag, and up  
to $\sim$3 mag in the most crowded regions, like b333).
In principle, PSF photometry is less accurate than aperture
photometry for bright, saturated sources. However, this effect
is mitigated in a field like b333 for two reasons: First, because 
of the extreme reddening, the bright and saturated sources are 
statistically less frequent. Second, the aperture photometry of bright sources
is performed by adopting large apertures, but the extreme crowding 
of b333 would hamper the results provided by this method.

Our photometric data set consists of 104 epochs in the $K_s$ band 
and two epochs in each of the $ZYJH$ bands. We consider the full
$K_s$-band time series and the average magnitudes in 
the other filters.

% Fig.~\ref{fig:magerr} shows a density plot of 
% the photometric error versus the measured magnitude
% in the $K_s$ band for the single phase points. $eK_s$
% is capped at 0.543 mag 
% 
% % read_cor01_stats.read_cor01('tile/b333/416_z_k.cor01','tile/b333/416_zjd_k','tile/b333/416_magerr')
% % cat *magerr > magerr_all (in realtà non le uso proprio tutti i paw...)
% % temp_180827
% 
% \begin{figure*}[!htbp]
% \centering
% \includegraphics[width=11cm]{mag_vs_err-eps-converted-to.pdf}
% \caption{}
% \label{fig:magerr}
% \end{figure*}

{\it Reddening} --- We adopted the 3D reddening map of
\citet[][henceforth S14]{schultheis2014}. This map provides a 6$\times$6 arcmin grid 
of E(\jmk) values in 21 bins of distance, from 0 to 10.5 kpc.
The reddening provided by this map depends not only on the coordinates but
also the distance of the target. Moreover we could not estimate the 
distance of all targets because not all of them are reliable distance
indicators (e.g., binaries, RV Tau (RVT) stars, non-pulsating variable stars). For 
these variables, we have adopted the 2D map by 
\citet[][henceforth G12]{gonzalez2012}, which is independent of distance.
The E($J-K_s$) values provided by this map are 
optimal for targets at the distance of the GC 
\citep[$\sim$8.3$\pm$0.2{[}stat.{]}$\pm$0.4{[}syst.{]} kpc,][]{degrijs2016}
but are overestimated for targets closer than the 
GC, and vice versa \citep{schultheis2014,braga2018b}.
To estimate the extinction $A_{ZYJHK_s}$ in the single bands, 
we adopted the reddening law of \citet{alonsogarcia2017}. We note that 
the G12 and S14 reddening maps and the adopted reddening law 
were obtained using VVV data. Finally, let us mention that for 
T2Cs, CCs, and anomalous Cepheids (ACs), 
we also derive independent reddening 
estimates by simultaneously solving  the PL$H$ and PL$K_s$ 
relations for distance and reddening.

{\it Proper motions} --- We adopt proper motions ($\mu_{l*}$,$\mu_b$) from the 
PSF photometry itself and $\mu_{l*}$ and $\mu_b$ were estimated with the same method used in 
\citet{contreras2017}. We note that these proper motions are relative
to the Galaxy. We tried to match the targets in our final list of 
candidate variable stars with the {\it Gaia}
DR2 \citep{gaia_alldr,gaia_dr2} source catalog, but we found reliable
matches for only 45 stars, of which only 30 have a five-parameter 
astrometric solution. Since these represent the minority of our sample of candidate periodic variable stars (PVSs)
and parallaxes are negative for almost half of the sample, indicating an
uncertain astrometric solution, we did not use these data. 
This was expected since {\it Gaia} performances
in extremely crowded regions are not optimal.

%temp_180821

\section{Analysis of the light curves}

We found 5,147,696 sources with $K_s$-band light curves within the tile. 
Such an amount of data requires proper selection to be manageable, and 
therefore we adopted rejection criteria at several steps
of the data analysis.

\subsection{Preliminary target selection}\label{selectiontarget}

\begin{figure*}[!htbp]
\centering
\includegraphics[width=10cm]{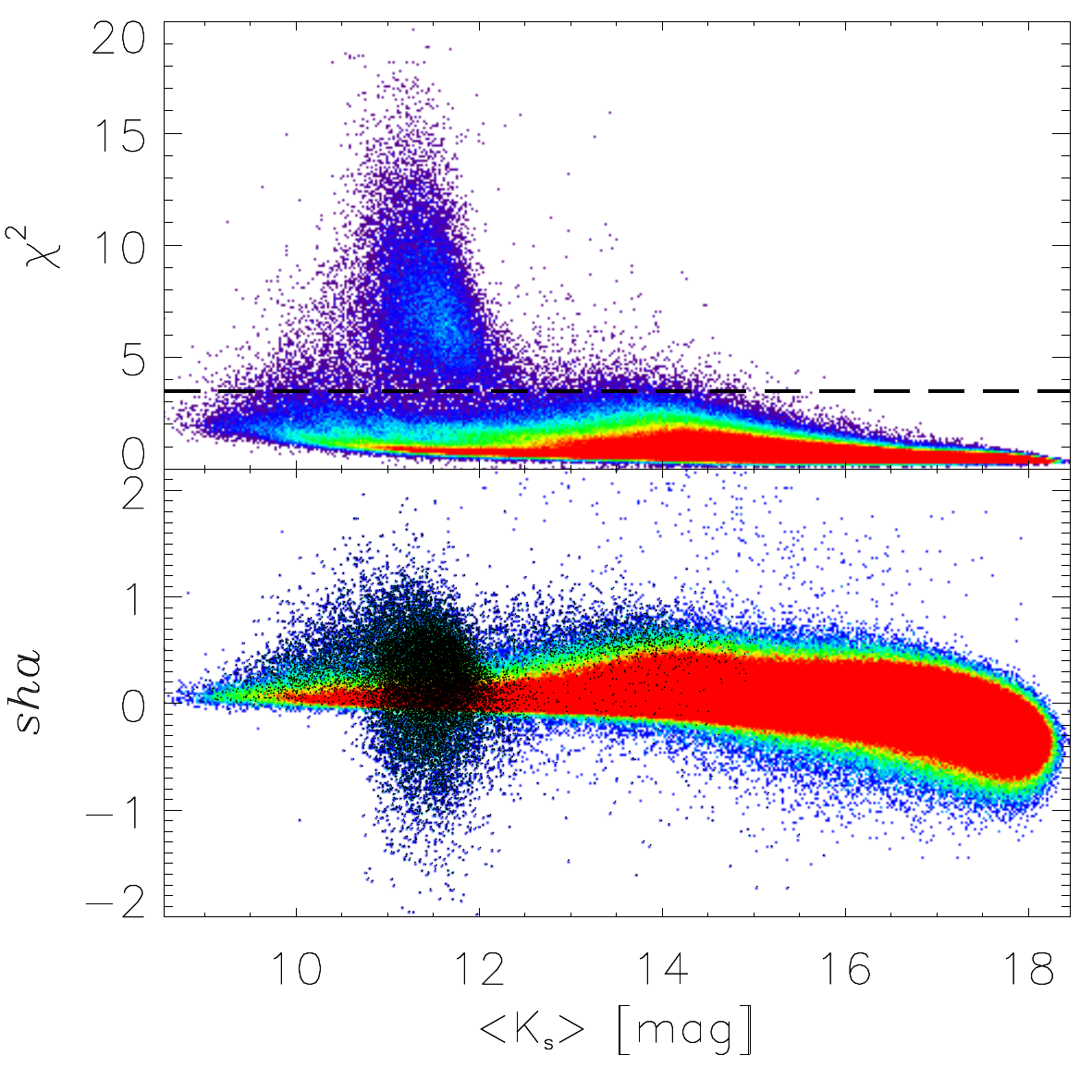}
\caption{${\chi}^2$ (top panel) and $sha$ (bottom panel) parameters
as a function of $K_s$-band magnitude. The density map goes from blue
(least dense regions of the plane) to red (densest). The thick dashed line in the 
top panel shows the ${\chi}^2$ cut applied, to reject all 
sources for which the photometric solution is poor. These sources are
marked as black points in the bottom panel.}
\label{fig:chi}
\end{figure*}

As a first step, we rejected all the sources that have light 
curves with either few phase points or poor-quality photometric reduction,
as well as those that are considered to be nonvariable within the photometic 
uncertainties. More specifically, we rejected those for which
a) the light curve has less than 25 phase points
with a valid PSF reduction output \citep{gran2015,navarromolina2019};
{ b)} the $\chi^2$ of the PSF fitting \citep{stetson94} 
is larger than 3.5. $\chi^2$ , which indicates the difference 
between the pixel counts in the CCD and the PSF that fits the 
source. We noted that the sources with $\chi^2 >$ 3.5 trace a 
``plume'' in the sharpness ($sha$)-magnitude plane (see 
Fig.~\ref{fig:chi}). The $sha$ parameter indicates whether the profile of 
the source is broader or narrower than the PSF profile, and 
is useful to detect nonstellar sources, which have 
large absolute values of $sha$;
{\it c)} the difference between the maximum and the minimum magnitude 
of the time series $\Delta K_s$ is
smaller than 0.1 mag. We note that this is not a cut in the $K_s$-band
amplitude ($Amp(K_s)$). In fact, due to the scatter of the phase points, 
$\Delta K_s$ is always larger than $Amp(K_s)$ and the quoted cut still allows 
us to detect variables with $Amp(K_s)$ down to $\sim$0.05 mag;
{\it d)} the median uncertainty on the 
phase points is larger than 0.2 mag.
We note that a magnitude uncertainty of 0.2 mag 
means a $\sim$20\% uncertainty on the flux; and
{\it e)} the variability indices based on even statistics 
\citep{ferreiralopes2016,ferreiralopes2017} 
indicate a nonvariable nature for the star.
We did not reject saturated \citep[$K_s<$12 mag,][]{contreras2017}
stars because, although mean magnitudes and amplitudes might
be inaccurate, periods are unaffected by saturation, at least
for variables with $Amp(K_s)\gtrsim$0.2 mag. Moreover, 
we are interested in matching our variable stars with Miras, 
CCs, and T2Cs found by \citet{matsunaga2009b,matsunaga2013}. Their
survey was shallower than the VVV due to the smaller diameter of their 
telescope (IRSF, 1.4m); therefore all their variables have 
mean $K_s$ magnitudes brighter than 14 mag and more than 
half are brighter than 12.5 mag.
We discuss  the offset in mean magnitude
due to saturation in more detail in Section~\ref{sect_caveat}.

% % Per il chi, devo per prima creare i files leggibili:
%   pawprints=['110', '112', '114', '116', '12', '14', '16', '18', '21', '211',
%          '213', '215', '23', '25', '27', '29', '310', '312', '314', '316',
%          '32', '34', '36', '38', '410', '412', '414', '416', '42', '44',
%          '46', '48', '51', '511', '513', '515', '53', '55', '57', '59',
%          '610', '612', '614', '616', '62', '64', '66', '68']
%    for paw in pawprints:
%       read_cor01_chisharp.read_cor01_chisharp('tile/b333/'+paw+'_z_k.cor01',
%           'tile/b333/'+paw+'_zjd_k','tile/b333/phased_lcv/'+paw+'/chisharp_'+paw)
% % poi lanciare temp_180822

The quoted criteria are driven by the requirements to have 
enough points to obtain a solid periodogram (point {\it a}), 
a good photometric solution for the accuracy of the 
measured magnitudes (point {\it b}), an amplitude which is 
large enough to detect variability over noise (point {\it c}),
and measured magnitudes which are more precise than 
20 \% (point {\it d}). In the following, we discuss  point {\it (e) }in more detail.

\citet{ferreiralopes2017} presented new dispersion and 
shape parameters for distributions with an even number 
of items in order to decipher whether or not the input 
distribution -- which must be an array of observed 
magnitudes -- could be representative of the light curve of 
a PVS. We note that only the 
measured magnitudes are taken into 
account, and not the epochs of observation.
We have adopted the seven parameters
$ED_{\sigma\mu}$, $ED_{\sigma m}$, $ED_{\mu}$, $ED_{m}$, 
$ED$, $ED_{(1)}$ and $ED_{(2)}$, as defined in
\citet[][Table 1]{ferreiralopes2017}. To avoid ambiguity, 
hereafter we refer to these seven parameters
as EVP$_i$, where $i$ ranges from 1 to 7.

\begin{figure*}[!htbp]
\centering
\includegraphics[width=7cm]{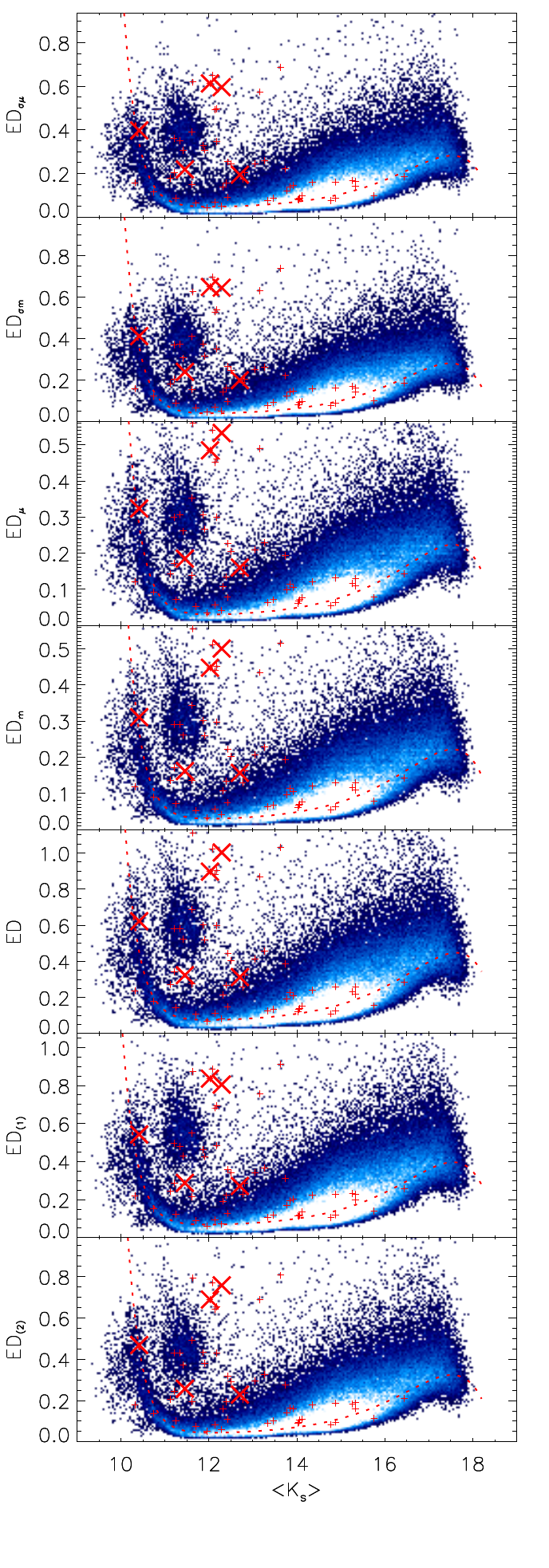}
\vspace{-1cm}
\caption{Density plots of the EVPs for the PCC 215. The density
scale goes from dark blue (low density) to white (high density).
The dashed lines represent the modified Strateva models. 
The red symbols display the candidate PVSs in this field of
view. Small red plus symbols represent candidate PVSs with ambiguous or
uncertain classification, and large red crosses represent 
candidate PVSs with reliable classification
(see Section~\ref{sect_classification}).}
\label{fig:strateva}
\end{figure*}

As a first step,  for each of the seven 
EVP$_i$ in each of the 48 PCCs, we derived the modified Strateva 
noise models \citep[][Eq. 18]{ferreiralopes2017} 
as a function of the mean magnitude. We adopted the modified models instead
of the classical ones \citep{strateva2001,sesar2007} because
the former better follow the distribution in the EVP$_i$-versus-magnitude plane, especially at 
bright magnitudes \citep[see Fig. 4 in][]{ferreiralopes2017}. 
We note that the modified Strateva models that we derived also correctly 
follow the distribution at faint magnitudes
(see Fig.~\ref{fig:strateva}, although \citet{ferreiralopes2017} 
pointed out that the modified Strateva models at faint magnitudes 
deviate from the distribution.

% per la figura: temp_180829

% controlla gli I_P delle candidate PVS a posteriori!!! temp_180914

For each of the EVP$_i$, we calculated $I_{(P)i}$, 
as defined in \citet[][Eq. 19]{ferreiralopes2017}. 
By construction,
$I_{(P)i}<1$ indicates a nonvariable star according to the $i$-th
EVP. One may also take account of the seven EVPs altogether and 
reject -- as nonPVSs -- all the targets for which the 
sum of the $I_{(P)i}$ ($\Sigma I_P$) is less than seven. W note that this
is possible because for a given target the seven 
$I_{(P)i}$ are all of the same order of magnitude. % temp_180910 per verificare quanto siano simili tra di loro gli EVPs
However, we have empirically
checked that there are some bona fide PVSs -- especially those
with small amplitudes -- for which $\Sigma I_P<7$ 
(see Fig.~\ref{fig:strateva} which shows that 
several candidate PVSs are below the $I_{(P)i}=1$ locus.). Therefore, 
we decided to adopt a more conservative approach and 
we kept, as candidate PVSs, all the targets for 
which $\Sigma I_P>3.3$. This threshold was chosen as a compromise between a 
moderate increase of false alarms and a higher completeness.

% xxx sokolovski 2016,2018

\subsection{Preliminary phase point selection}\label{selectionpoints}

After the rejection based on the criteria $a)$, $b)$, $c)$, 
$d)$ and $e)$ in Section~\ref{selectiontarget}, 
where $\sim$15\% of the sources were discarded, 
 for each light curve we performed a 
rejection of the phase points that we assumed to be  
outliers. The rejection process was based on
an iterative sigma clipping at a 3.5$\sigma$
level with respect to the median magnitude. This is a conservative threshold that 
allowed us to detect periodic variable stars with large amplitudes 
(up to $\sim$2 mag) and at the same time to discard outlying 
phase points that could affect the accuracy of the periodicity search. 
In this step, for $\sim$57\% of the light curves, we do not reject any
phase points. For $\sim$23\%, $\sim$10\%, $\sim$5\%, and $\sim$2\% 
of the light curves, the number of points that were rejected was one,
two, three, and four, respectively. For the remaining $\sim$3\% of
the light curves, five or more points (up to eighteen) were rejected.

\subsection{Periodicity search}\label{sect_periodsearch}

After the selections and cuts described in the above section, we ran our periodicity 
search algorithm over all the light curves. As a preliminary step, we 
converted our epochs from Julian dates into heliocentric Julian dates ($t_{HJD}$). 
We calculated both the classical Lomb-Scargle \citep[LS;][]{scargle82}
and the Generalized Lomb-Scargle \citep[GLS;][]{zechmeister2009} periodograms.
The difference between the two is that the zero point of the model function
in the LS is fixed at the empirical mean magnitude of the phase points, while
in the GLS method, the zero point is allowed to float. Therefore, the GLS 
is more reliable when the light curve is not well sampled 
\citep{vanderplasivezic2015}. 

We performed both a LS and a GLS analysis because the 
comparison of the periodograms is useful to detect and 
automatically reject stars that are not periodic variables.
We have verified empirically that, if the highest peak of the GLS 
periodogram does not match -- within the typical peak
width -- any among the ten highest peaks 
of the LS periodogram, the peaks are associated with light 
variations that are not strictly
periodical, and therefore we can reject the 
target as a nonPVS. In the following sections, we 
explain the details of our analysis.

\subsubsection{Frequency grid}\label{sect_freqgrid}

To calculate the periodograms, we used an evenly spaced 
grid of frequencies, from $\nu_{min}$=$\dfrac{1}{\Delta t}$ days$^{-1}$ -- where
$\Delta t$ is the total time between the first and the last epoch of the light curve in days -- to 
$\nu_{max}$=0.995 days$^{-1}$. This range was chosen based on the following considerations.
{\it 1)} Our first aim is to detect T2Cs and fundamental-mode (FU) CCs, which have periods from
$\sim$1 to $\sim$200 days. {\it 2)} We did not include frequencies that are too close to 1 
day$^{-1}$ because the increase in the number of genuine variables would come at the 
cost of a much higher increase of aliases around 1 day.
Based on the typical number of phase points for our light curves, the limit at 0.995 
days$^{-1}$ allows us to automatically reject most of the aliases. 
{\it 3)} We did not extend $\nu_{max}$ to the 
range of frequencies of RRLs because the computation 
time would be around ten times longer with $\nu_{max}=5$ days$^{-1}$.
Although the limit at $\nu_{max}$ might generate aliases, we 
adopted an a posteriori LS analysis on a much smaller sample to 
solve the aliases (see Section~\ref{sect_refining}).

% 3) We have selected a 
% frequency range which covers periods
% up to 1000 days because this comes at a negligible
% computation time cost (less than 1\% more time is needed to increase the period 
% limit from 200 to 1000 days), but allows the detection of 
% Miras and other Long-Period Variables. A lower frequency limit would be useless
% because our VVV data span a maximum of $\sim$1861 days, from the first to the last 
% epoch, and we need at least two cycles to detect regular (or semi-regular) PVSs.

The number of points in the frequency grid of 
the periodogram ($N_{freq}$) depends on the 
range of epochs covered by the light curve
($\Delta t$). As a matter of fact, $Df = \Delta t^{-1}$
is the characteristic width of a peak in the periodogram. Therefore, we set $N_{freq} = int(n_{spp}\cdot \Delta t \cdot(\nu_{max}-\nu_{min}))$, 
where $n_{spp}$ is the oversampling factor \citep{graham2013} and 
$n_{spp}$ corresponds to a phase shift 
of about 0.1 between neighboring frequencies that ensure the signal 
detection of almost all variable types \citep[for more detail see ][]{ferreiralopes2018}.
We set $n_{spp}=9$ for all the light curves
\citep[normally, the recommended value of $n_{spp}$ is between 5 and 10;][]{vanderplasivezic2015}.
Since $\Delta t$ is not the same for all the light curves, 
$N_{freq}$ ranges between $\sim$3,000 and $\sim$16,600, with a 
mean of 16,082.

\subsubsection{Height of the peaks and frequency aliases}\label{heightpeaks}

After obtaining the periodograms, 
we empirically verified that if the peaks are not
high enough or if there are too many low peaks -- see details
below -- then the light curve is too noisy to properly 
classify our targets.

Therefore, we perform a further selection 
of targets based on the intensity of the peaks of the periodograms. 
Adopting the standard deviation of the periodogram itself ($\sigma_P$), 
we discarded 
1) light curves with less than 30 points and no peaks 
higher than 8$\sigma_P$;
2) light curves with 30 points or more, and no peaks 
higher than 9$\sigma_P$; and
3) light curves with more than 140 peaks higher than 5$\sigma_P$ 
and no peaks higher than 12$\sigma_P$.

After this further selection, we adopted a procedure to 
check---and eventually correct---aliased frequencies. Following
\citet{vanderplasivezic2015} and \citet{vanderplas2018}, we have 
checked whether the frequency of the highest peak
($\nu_{hi}$) is either a harmonic
of the fundamental frequency, an alias, or the true 
fundamental frequency. This is done in two steps.
First, we checked whether or not there were sufficiently high peaks (larger than 
6 or 7$\sigma_P$ if the number of phase points is, respectively, smaller or 
larger than 30) around the critical frequencies ($\nu_{check}$).
To check for harmonics, $\nu_{check} = \dfrac{\nu_{hi}}{m}$, with $m$=2,3; 
to check for aliases, $\nu_{check}=|\nu_{hi}+n \cdot \nu_{alias}|$, 
where $\nu_{alias}$=1 day$^{-1}$ and $n$=$\pm1, \pm2$.
Second, we compared the $\chi^2$ of the Fourier fits at $\nu_{hi}$
and $\nu_{check}$, and selected the best frequency as that 
which minimizes the $\chi^2$.

Finally, we obtained best frequency estimates for 889,663
targets, of which 435,126 and 454,537 are from GLS and LS 
periodograms, respectively. 

We folded the light curves 
calculating the phases $\phi$ as the decimal 
part of $\dfrac{t_{HJD}-T_0}{P_{best}}$,
where $T_0$ is an arbitrary zero epoch that we set at 0.0 days.

\subsection{Light-curve fit}

Despite the quoted cuts and selections, after the analysis
of the periodograms, we still have almost one fifth from the 
original sample. Therefore, further selections are needed to
narrow down the final sample to PVSs only.

For this purpose and to derive the pulsation properties of the 
targets (mean magnitude, amplitude, and uncertainties), 
we fitted the folded light curves with 
a Fourier series of the second order ($F(\phi)$), and used the 
properties of the fit to select our candidate
variable stars. We use the second order because, at this stage, 
the sample of light curves is still not completely free from
nonPVSs, and Fourier series of higher 
order would provide unreliable fits with unphysical bumps.

\subsubsection{Mean magnitudes and amplitudes}

After obtaining the Fourier fit of the light curves, we derived the 
mean magnitudes $\langle K_s \rangle$ as the integral of the fits converted
to flux, and the amplitudes $Amp(K_s)$ as the difference between 
the brightest and the faintest points of the fits.
We estimated the uncertainties on the mean magnitude ($eK_s$)
as the sum in quadrature of the standard deviation of the phase 
points around the fit plus the median photometric 
error on the phase points. The uncertainty on the amplitude $eAmp(K_s)$
was derived as the sum in quadrature of the median photometric errors 
of the phase points around the maximum and minimum, plus the standard deviation of
these phase points around the fit of the light curve. The final
value was weighted with the number of phase points around the 
maximum and the minimum \citep{braga2018}. The 
fits might show, at most, two secondary minima and 
maxima. In this case, we estimate the amplitude of 
the bump $Amp(K_{s(bump)})$ as the difference between the 
secondary minimum and maximum.

\subsubsection{Selection of the final sample}

With almost 900,000 targets remaining, we applied 
further selections aimed to narrow down the sample 
to a more manageable size.

{\it 1) Selection on phase gaps} --- A fraction of 
folded light curves display wide gaps in phase 
($\geq$0.25 cycles). We empirically checked that 
for stars with $P_{best} \leq 330$ days, these gaps 
are caused by a poor estimate of $P_{best}$ and that 
the target is not a PVS. Therefore,
we reject all targets with a phase gap wider than 
0.25 cycles and $P_{best} \leq 330$ days. Targets
with wide gaps but longer periods are kept in the sample
of candidate PVSs because bona fide PVSs with long 
periods  also show such gaps, caused by the 1 year alias.

{\it 2) Selection on amplitude} --- We put an upper limit of 
$Amp(K_s) \leq$3.5 mag on the light amplitude. This is a very conservative
upper limit for the light amplitudes of Miras, which are the 
variables with the largest amplitudes among those that
we are interested in. This threshold  
is based on photometric surveys of Miras in the Bulge, both in 
the NIR \citep[][which does not find any Mira with 
$Amp(K_s) \gtsim$2.7 mag]{matsunaga2009b}, and in the optical
(\citealt{soszynski2013}, which does not find any Mira with 
$Amp(I) \gtsim$7 mag; we note that for Miras we can assume a 
ratio $Amp(K_s)$/$Amp(I)\approx$ 0.45, \citealt{whitelock2012}).
Moreover, the accuracy and precision of
our data do not allow to detect PVSs with $Amp(K_s)<0.03$ mag
\citep{gran2015}. Therefore, we only keep the targets 
with 0.03 $\leq Amp(K_s) \leq$ 3.5 mag.

% {\it 3) Selection on bumps} --- If 
% $\dfrac{AK_{s(bump)}}{Amp(K_s)}>0.15$---which means that 
% the fit displays a deep bump---we discard the target 
% because neither T2Cs, nor any of the other types of variables of
% the instability strip, display this feature in the light curve. This kind
% of deep bumps usually shows up from the fits of either 
% too noisy or non-PVS light curves.
{\it 3) Selection on bumps} --- Second-order Fourier fits might 
show local minima and maxima. Although some variables of the IS do 
show these features, they are not particularly deep, especially at
long wavelenghts \citep{laney1993}. Therefore, if 
$\dfrac{Amp(K_{s(bump)})}{Amp(K_s)}>0.15,$ which means that 
the fit displays a deep bump, we discard the target because
these features in the light curve fit show up in 
noisy or nonperiodic light curves.

\begin{figure*}[!htbp]
\centering
\includegraphics[width=7cm]{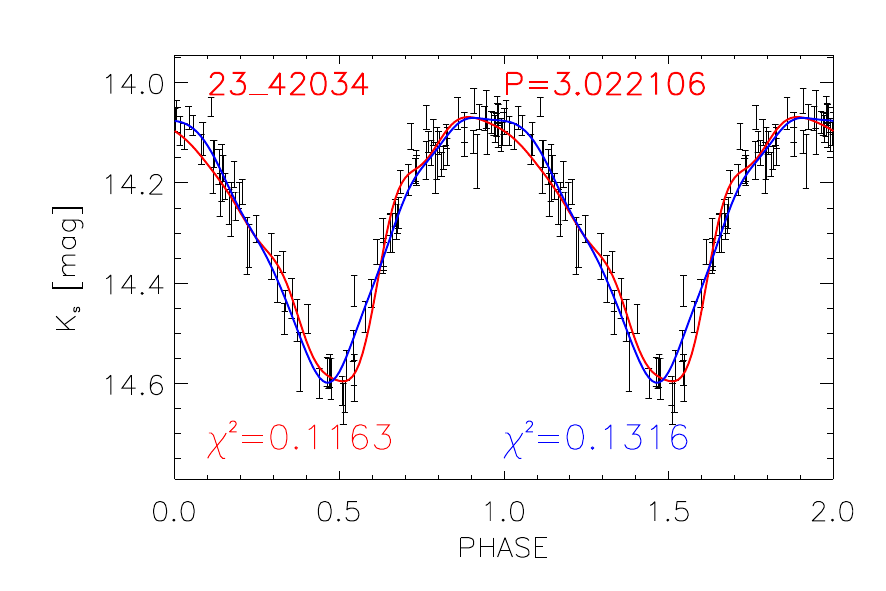}
\caption{Light curve of the variable star b333\textunderscore23\textunderscore42034, a T2C. 
The name and the period (in days) are labeled on the top. 
The red and blue lines display the CC and the T2C template fits, 
respectively. The ${\chi}^2$ of the two fits are labeled at
the bottom. The colors of the labels are the same as those of the fits.}
\label{fig:lcv_template}
\end{figure*}

{\it 4) Selection on template fit} --- Templates of 
NIR ($JHK_s$) light curves of T2Cs are available
\citep{bhardwaj17b}. For all targets with 
1 day $\leq P_{best} \leq$ 80 days, we derived 
the template fit of the light curve. We note that we 
did not properly apply the template procedure to 
find the mean magnitude because this would 
require the knowledge of the epoch of maximum light. 
We performed a least-squares minimization of the 
template function $T(\phi)$ with respect to three independent 
variables, which are the shift in magnitude 
($\Delta mag$), the shift in phase ($\Delta \phi$), and 
the amplitude of the template fit $AK_{s[T(\phi)]}$.
We note that in this process the shape of the template 
fit is fixed. Therefore, the least-squares minimization process
only shifts the fitting function horizontally, vertically, 
or stretches its amplitude. We subsequently compare the template
fit and the Fourier fit. If they are similar, this means 
that the target can be considered as a candidate T2C. Therefore, 
we compute a parameter ($c_{10}$) to quantitatively compare the two 
fits, defined as the average of the absolute value of the 
difference between the two fits, divided by the average 
amplitude of the two fits: 
$c_{10}=\dfrac{\langle|F(\phi)-T(\phi)|\rangle}{(AK_{s[F(\phi)]}+AK_{s[T(\phi)]})/2}$. 
% XXX figura con una scartata e una buona
If $c_{10}>0.2$, the two fits are quite different 
from one another, therefore we can discard the target from
the list of candidate PVSs. We note that 0.2 is a 
conservative threshold that allows for targets that 
are plainly not PVSs to be rejected, but also allows us to keep eclipsing binaries
(EBs) with light curves that are quite different from those of T2Cs.
Taking account of the accuracy of our data, 
we can assume that the shape of the light curves of CCs is
similar to those of T2Cs. We tried to
fit the light curves using also the templates 
of CCs \citep{inno15}, trying to separate 
between T2Cs and CCs, based on the ${\chi}^2$ of 
the two fits. However, this was not possible
because the two templates are very similar over the 
whole range of periods. As a matter of fact, for the majority of our 
light curves (all those with $\langle K_s \rangle \gtsim$
14.5 mag) the ${\chi}^2$ of the two fits are often similar. 
Figure \ref{fig:lcv_template} shows the light curve of a variable which is clearly 
a T2C (based on both its location in the Bailey diagram and 
its proper motion) despite the $\chi^2$ of the T2C template
fit being larger than that of the CC template fit (see labels).

% {\it 5) Selection on secondary parameters} --- Without going 
% into detail, we defined secondary parameters that we used 
% to reject targets which are unlikely to be PVSs. We took 
% account of the number of phase points, of the amplitudes and of the 
% photometric uncertainties of the phase points to put constraints 
% and further reduce the sample. As a whole, the effect of these
% selections is to reject XXX of the targets remaining before
% the Fourier fitting.

\begin{figure*}[!htbp]
\centering
\includegraphics[width=7cm]{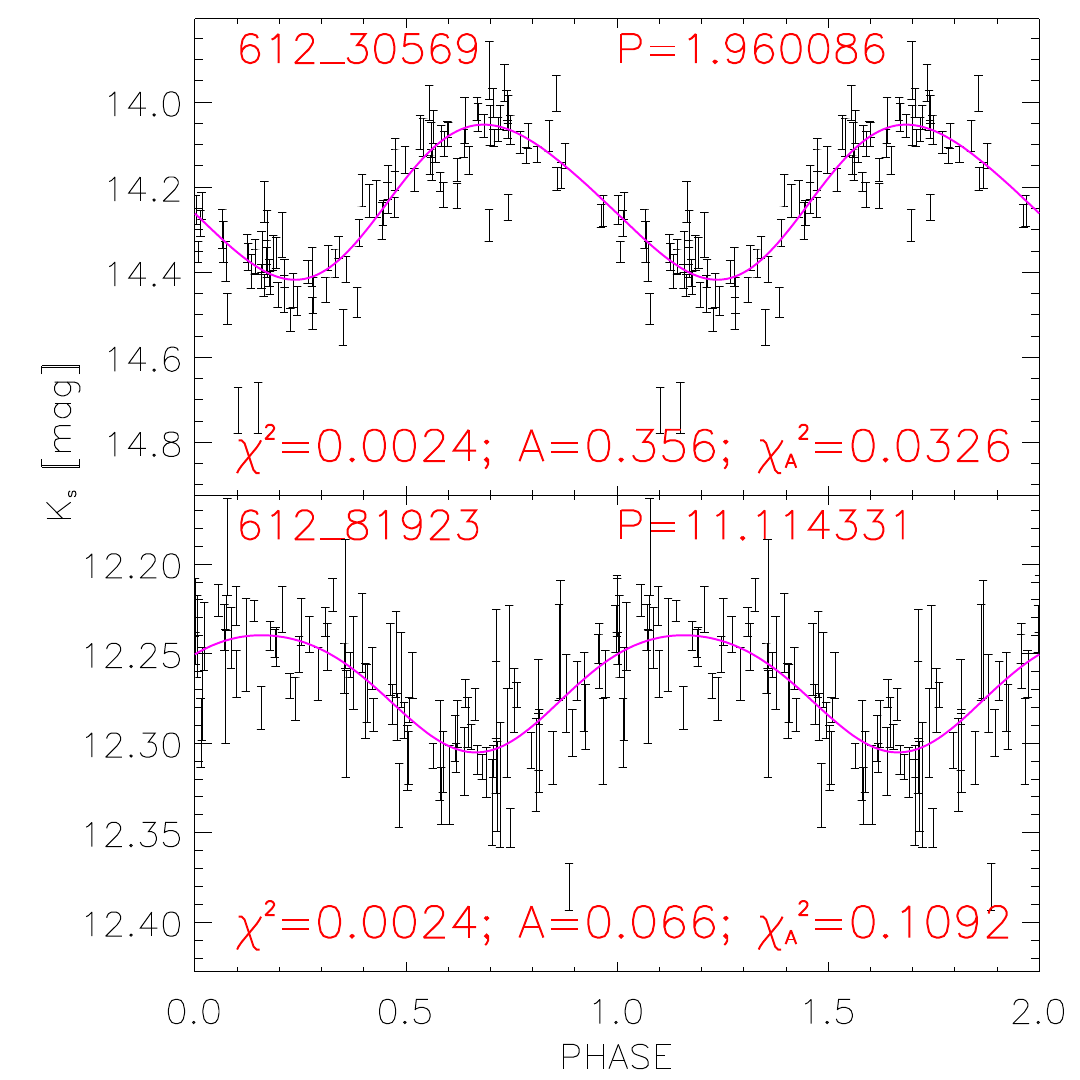}
\caption{Top: Light curve of the variable \textunderscore612\textunderscore30569.
The name and period (in days) are labeled at the top. The Fourier fit is 
displayed as a magenta line. The ${\chi}^2$, $Amp(K_s)$ and ${\chi_A}^2$
are labeled at the bottom. Bottom: As in the top panel, but for \textunderscore612\textunderscore81923.}
\label{fig:lcv_chisq}
\end{figure*}

{\it 5) Selection and ranking on modified ${\chi}^2$} --- 
A further selection is based on the dispersion 
of the phase points around the 
fit. However, we did not compute a simple 
${\chi}^2 = \dfrac{{\Sigma}_i (mag_i-F(\phi_i))^2}{n}$, 
where $mag_i$ is the measured magnitude of the $i$-th phase 
point and $F(\phi_i)$ is the value of the Fourier
fit at the phase $\phi_i$ of the $i$-th phase point.
We adopted a modified chi squared 
${\chi_A}^2 = {\chi}^2/{Amp(K_s)}^2$ that we
also used  to rank our final sample of targets for 
the visual inspection, from the ``clearest'' 
(smaller ${\chi_A}^2$) to the ``least clear light'' curves.
The division by the squared amplitude has the 
effect of increasing the ${\chi_A}^2$ of the small-amplitude 
variables and in turn  decreasing their rank. This 
operation is needed to obtain a more reliable ranking:
Figure~\ref{fig:lcv_chisq} shows two variables that have an
identical ${\chi}^2$ but different $Amp(K_s)$. Based on ${\chi_A}^2$,
the variable in the upper panel is ranked higher than the one in
the bottom panel to take account of the relative dispersion with
respect to the amplitude. Finally, we rejected all the targets for 
which ${\chi_A}^2 > 0.2$.

% minchisq2_all in searchvariables_pawprint.pro
% che viene scritto in folder+/paw/fitcoeff_*
% uso temp_180911.pro per cercare variabili con chisq simile e ampl diversa

{\it 6) Visual inspection} --- After the quoted   
steps, we were left with 54,667 targets, which is a little more
than 1\% of the starting sample.
This is still too large and contains many stars that cannot be considered 
as bona fide candidate PVSs. We performed a visual inspection and 
found 1,013 ``top tier'' candidates. We note that the final 
catalog includes 1,019 targets because we manually added 
six Cepheids already found in the literature \citep{matsunaga11,matsunaga2013}
which are present in our initial target list but were not retrieved
by the periodicity search algorithm (see Section~\ref{sect_classification}
and ~\ref{sect_matsu13}).

\subsubsection{Refining the periods and fits}\label{sect_refining}

After the quoted selections it is possible to 
visually inspect the light curves of all the candidate variables 
to improve our analysis. As a matter of fact, these steps are mandatory for a 
more accurate classification of the PVS candidates.

{\it 1) Unresolved aliases} --- The alias check process performed
in Section~\ref{heightpeaks} has one disadvantage: it does not 
detect aliases at frequencies outside the frequency
grid (in our case, for $\nu>0.995$ days$^{-1}$). 
Therefore, we repeated the whole periodicity 
search process with LS and GLS on the 1,019 PVS candidates, but this time
on a frequency grid going from $\nu_{min}$=0.000 days$^{-1}$ to 
$\nu_{max}$=5 days$^{-1}$. Since the size of this sample is 
much smaller than the starting list of sources, this 
is not a time-consuming process. For 77 targets, we 
found a $P_{best}$ shorter than 1 day, which is  
different from that obtained in the first iteration.

{\it 2) Order of the Fourier fit} --- Some variables -- especially
the brightest ones, which have small photometric errors -- show 
well-defined light curves with low noise, and are better fitted by 
Fourier series of an order higher than the second. For these 
variables, we  repeated the fit using a
third-, fourth-, or fifth-order Fourier series. We also repeated the 
fits of almost sinusoidal light curves using a first-order Fourier fit to avoid unphysical bumps appearing 
especially in light curves with few phase points.

{\it 3) Frequency doubling} --- EBs, RVTs, but also WVs with periods
down to $\sim$16 days \citep{soszynski2017} might have light 
curves with alternating deep and shallow 
minima of light. For these targets, our algorithm 
for the periodicity search incorrectly assigned 
$P_{best}$ to the time interval between 
two adjacent minima. However, this is half of the true, 
physical variability period. To detect the alternating minima
and to better classify the variables, 
we visually inspected the light curves of all the PVS candidates
at both $P_{best}$ and 2$\cdot P_{best}$. We changed 
$P_{best}$ into 2$\cdot P_{best}$ for those variables with 
clear signs of alternating minima.

% WVir with alternating minima..... intermediate age???

\section{Classification}\label{sect_classification}

The stars in our final sample have 
periods between $\sim$0.38 and $\sim$1240 days. 
In principle, they could be of any type among the 
pulsating stars, that is, RRLs, T2Cs, CCs, ACs, long-period variables  
\citep[LPVs, which include Miras and SRVs, but not OSARGs, 
which have overly small amplitudes;][]{wray2004}, non-pulsating variables (NPVs), which include, for example,
EBs of any kind (detached, semi-detached,
W UMa contact), and spotted stars, that is, 
stars which show variability due to the interplay of
their rotation and the presence of large spots on their surface.
Normally, these objects are associated to the pre-main sequence population. 
To discriminate between these types of variables, 
we adopted the criteria listed below. 
We point out that none of the following criteria were adopted alone 
for the classification, but they were all considered together.
We list them in order of decreasing reliability.

{\it 1) Periods} --- Based on the literature, we adopted 
period thresholds for the quoted types of variables (see Table \ref{tab:vartypes}).
We adopted conservative limits because the period thresholds
of pulsating stars depend on many factors (metallicity, helium 
abundance, $\alpha$ enhancement) and a detailed 
summary is beyond the aim of this paper.
{\it 1a) NPVs} --- These variables cover a wide range of 
periods, which includes all the periods of our targets. 
{\it 1b) RRLs} --- Using OGLE data, 
\citet{soszynski14} found more than 38,000 RRLs, which is the largest  
homogeneous available catalog of RRLs. The period ranges of 
RRc and RRab are [0.20-0.54 days] and [0.28-1.00 days], 
respectively. These are quite usual 
thresholds, apart from the lower limit of RRab, for which
$\sim$0.4 is a more common value. However, since the shortest
period in our list of PVSs is $\sim$0.38 days, 
there is no difference in adopting one or the other value for the threshold.
{\it 1c) ACs} --- The  most extensive catalog of bona fide 
ACs is provided by the OGLE survey \citet{soszynski2015a}, for targets
in the Large Magellanic Cloud (LMC) and Small 
Magellanic Cloud (SMC). These latter authors found ACs
in a period range which is wider (0.38-2.7 days) than those  
quoted in theoretical studies 
\citep[$\sim$0.5-2 days][]{fiorentino06}. We point out that 
the GC is an environment which in principle does 
not favor the formation of ACs, which arise either from very
metal-poor or binary progenitors \citep{fiorentino12c}. 
Nonetheless, to date, 20 of them were found in the 
Bulge \citep{soszynski2017}. Therefore, 
we cannot discard a priori the possibility of finding ACs. 
{\it 1d) CCs} --- The sample of CCs with the widest range in 
period (0.25-208 days) is that of LMC and SMC CCs by OGLE 
\citep{soszynski08b,soszynski10a}. We point out that the adopted 
empirical ranges are wider than theoretical predictions 
\citep{bono00a}.
{\it 1e) T2Cs} --- The commonly accepted lower limit 
of the periods of T2Cs -- which marks their 
separation from RRLs -- is 1 day 
\citep{dicriscienzo07}. To set an upper threshold 
for T2Cs is a more delicate issue. RV Tau stars are the longest-period 
T2Cs and might show alternating deep and shallow 
minima. Therefore, two periods can be defined: the 
time passing between two deep minima (true period)
and the time between two adjacent minima (formal period, 
by definition, half of the fundamental period). 
According to \citet{wallerstein2002}, the maximum formal period 
is theoretically $\sim$75 days. On the observational side, the longest 
formal periods in T2C catalogs are shorter than 90 days for 
Galactic globular clusters \citep[][IRSF]{matsunaga06} and shorter than 80 days for the 
LMC, SMC, and Bulge \citep[][OGLE]{soszynski2017,soszynski2018}. 
However, it is not easy to detect the alternating minima, 
therefore it is not straightforward to discriminate between the 
two periods. Finally, we adopted a conservative threshold of 
100 days for the formal period.
{\it 1) Miras} --- These variables are often classified 
based on their amplitude ($A_I$>0.8 mag) rather than their 
period. The most common lower period limit in the literature
is 100 days \citep{itamatsunaga2011,catchpole2016}. Nonetheless, 
Miras with shorter periods -- down to $\sim$80 days -- were 
found in the Galactic Bulge \citep{soszynski2013} 
and furthermore \citet{whitelock2012} states that Miras could 
have periods shorter than 100 days. Therefore, 
we adopt 80 days as our threshold. We note that this is also
the lower limit given in the General Catalog of Variable Stars
\citep{samus2017} and the International Variable Stars Index 
definition of the class. On the other hand, the
periods of Miras can be as long as 2000 days 
\citet{whitelock2012}, which is  
longer than the longest period in our final sample of 
candidates.

\begin{table*}
 \footnotesize
 \caption{Table of criteria for the classification of variable stars.}
 \centering
 \begin{tabular}{l c c l c c l}
 \hline\hline  
 Type & $P_{min}$ & $P_{max}$ & Ref.\tablefootmark{a} & PL$H$ & PL$K_s$ &  Ref.\tablefootmark{b} \\
 & days & days & mag & mag & \\
\hline
NPSs     & <0.38     & >1000     & 1     & \ldots                                       &    \ldots                                    & \ldots \\ 
RRc(FO)  & <0.38     &$\sim$0.54 & 2     & \ldots                                       & --1.56--2.72$\cdot\log P$                    & 11     \\
RRab(FU) &$\sim$0.38 &$\sim$1    & 2     & --0.971--2.226$\cdot\log P$                  & --0.998--2.250$\cdot\log P$                  & 11     \\
ACFO     &$\sim$0.38 &$\sim$1.2  & 3     & \ldots                                       & --2.42--4.18$\cdot\log P$                    & 12     \\
ACFU     &$\sim$0.50 &$\sim$2.7  & 3     & \ldots                                       & --1.74--3.54$\cdot\log P$                    & 12     \\
CCFO     & <0.38     &$\sim$6.3  & 4,5   & --2.851--3.455$\cdot\log P$                  & --2.890--3.455$\cdot\log P$                  & 13     \\                 
CCFU     &$\sim$0.80 &$\sim$210  & 4,5   & --2.366--3.227$\cdot\log P$                  & --2.408--3.245$\cdot\log P$                  & 13     \\
T2C      &$\sim$1    &$\sim$100  & 6,7,8 & --3.358--2.202$\cdot\log P$\tablefootmark{c} & --3.418--2.232$\cdot\log P$\tablefootmark{c} & 14     \\
Miras    &$\sim$80   &$\sim$2000 & 9,10  & \ldots                                       & --6.865--3.555$\cdot\log P$\tablefootmark{d} & 15     \\
\hline
 \end{tabular}
 \tablefoot{ 
 \tablefoottext{a}{1: \citet{soszynski2016b}, 2: \citet{soszynski14}, 3: \citet{soszynski08c}, 
 4: \citet{soszynski08b}, 5: \citet{soszynski10a}, 6: \citet{matsunaga06},
 7: \citet{soszynski2017}, 8: \citet{soszynski2018}, 9: \citet{whitelock2012},
 10: \citet{soszynski2013}.} 
 \tablefoottext{b}{ 11: \citet[][conversion to 
 VISTA photometric system negligible, because the coefficients have 
 only two significant decimal digits]{marconi15}, 12: \citet[][conversion to 
 VISTA photometric system negligible, because the coefficients have 
 only two significant decimal digits]{ripepi2014}, 13: \citet[][converted to 
 VISTA photometric system]{inno2016}, 14: \citet[][converted to 
 VISTA photometric system]{bhardwaj17b}, 15: \citet[][converted to 
 VISTA photometric system]{matsunaga2009b}. The zero points of the 
 relations based on LMC variables were derived assuming, as the distance
 of the LMC, $d_{LMC}$=49.59$\pm$0.09$\pm$0.54 kpc \citep{pietrzynski2019}} 
 \tablefoottext{c}{ We did not derive distances of candidate T2Cs 
 with $P_{best}>$20 days (RVTs) because there is no general consensus on
 whether RVTs can be used as distance tracers or not 
 \citep{matsunaga06,ripepi2015,bhardwaj17b,braga2018b}.}
 \tablefoottext{d}{ C-rich Miras with periods longer than 
 $\sim$320-350 days can be significantly affected by circumstellar reddening
 and do not obey the PL relation \citep{matsunaga2009b,itamatsunaga2011}.
 Since our color estimates are not accurate enough 
 to separate C-rich and O-rich stars, we conclude that distance
 estimates for long-period Miras are not reliable.}}
 \label{tab:vartypes}
\end{table*}

{\it 2) Bailey diagram} --- We have collected the periods and $Amp(K_s)$ 
of RRLs, T2Cs, CCs, Miras and ACs in the literature, in the 
field of the Milky Way, the LMC, the SMC, the Galactic Globular Clusters and the Galactic Bulge 
\citep{matsunaga06,matsunaga2013,ripepi2014,inno15,ripepi2015,gran2016,matsunaga2016,ripepi2016,yuan2017}.
For each candidate PVS, we have checked its position in the Bailey diagram and 
compared this with the position of variables of known type, to narrow 
down the classification.

{\it 3) Distance and position} --- Based on known calibrations 
of PL relations for the different types of variables
(see Table~\ref{tab:vartypes}),  for 
 each target, we have calculated a set of provisional distances:
$d_{RRab}$, $d_{RRc}$, 
$d_{CCFO}$, $d_{CCFU}$, $d_{ACFO}$, $d_{ACFU}$, $d_{T2C}$, $d_{Mira}$. 
These must be interpreted as the distance at which the target 
would be located if it was a variable of a given type, as 
in the subscript. Using the ($l$,$b$) coordinates, we also 
derive the provisional positions in Cartesian coordinates---$x$, 
$y$ and $z$---of the targets in the Galaxy. We note that there are several caveats on distance.
{\it 3a) Metallicity dependence} --- At NIR wavelenghts, 
the effect of metallicity on the PL
relation of CCs is reduced compared to the optical 
\citep{marconi10,bhardwaj15}. The effect of metallicity 
is negligible also on the PLs of T2Cs \citep{dicriscienzo07} 
and Miras \citep{whitelock2008}. There is no solid theoretical or empirical 
evidence of metallicity dependence of the PL$K_s$ of 
ACs \citep{fiorentino06} either. On the other hand, the zero-point of 
the PL$K_s$ relation of RRLs is affected by metallicity
\citep{bono01,catelan04,marconi15}. We assume 
[Fe/H]=--1.0 \citep{sansfuentes2014,hajdu2018} as a
typical iron abundance of Bulge RRLs. We note that the error propagation
associated to the uncertainty on [Fe/H] is negligible compared to 
the other factors. 
{\it 3b) RVTs ---} There is currently debate 
over whether RVTs follow the PL relation of T2Cs 
\citep{matsunaga06,bhardwaj17b} or not 
\citep{wallerstein2002,ripepi2015}. More specifically, 
it is not clear how to separate intermediate-age RVTs
from old, low-mass RVTs, which have a completely different
evolution; it might be appropriate to give them a different name 
\citep[e.g., V2342 Sgr stars,][]{catelan15}. To sum up, 
the distances of even bona fide RVTs cannot be trusted.
{\it 3c)\ Reddening} --- We are aware that the pixels of the S14 reddening
map are too large to provide accurate E($J-K_s$) estimates of targets within 
a region where the reddening pattern is so irregular at small scales. In 
Section~\ref{sect_caveat}, we discuss in more detail the comparison 
of reddening and distance with other works. 
We point out that for a fraction 
of targets our distance estimates might be overestimated due 
to underestimated extinction. However, NIR multi-band
light curve templates are available for RRLs, CCs, and T2Cs  
(\citealt{braga2019}, \citealt{inno15} and 
\citealt{bhardwaj17b}, respectively), providing
the possibility to estimate accurate $J$- and $H$-band mean 
magnitudes ($\langle J \rangle$, $\langle H \rangle$) and, in turn,
to derive independent distance and $A_{Ks}$ estimates. 
This is the same approach used by 
M09 and M13, who did not use any reddening map.
We point out that although we did derive $\langle J \rangle$, 
we only employed PL$K_s$ and PL$H$ relations to estimate
extinction and distances, because differential reddening
and variations of the reddening law have an overly large effect
on targets in this region. The net effect of employing the
PL$J$ relations is to increase distances by $\sim$0.4 kpcs, 
leading to a very unlikely peak of T2C distances around 
$\sim$8.9 kpc.

To estimate the $\langle J \rangle$ and 
$\langle H \rangle$ of RRLs and CCs, 
we used the light-curve templates of RRLs 
\citep{braga2019} and CCs \citep{inno15}. For T2Cs, 
only the $K_s$-band light-curve templates 
are available. However, they provide accurate mean magnitudes also
when applied to the $H$-band data \citep{bhardwaj17b}, since 
the light-curve morphology in these two bands is very similar. 
Finally, we adopted the $J$-band CCFO light-curve
template by \citep{inno15} to derive $\langle J \rangle$
for T2Cs. We checked that, within the photometric
errors,  no difference is seen when using the phase
correction \citep{bhardwaj17c}.

\begin{figure*}[!htbp]
\centering
\includegraphics[height=6.5cm]{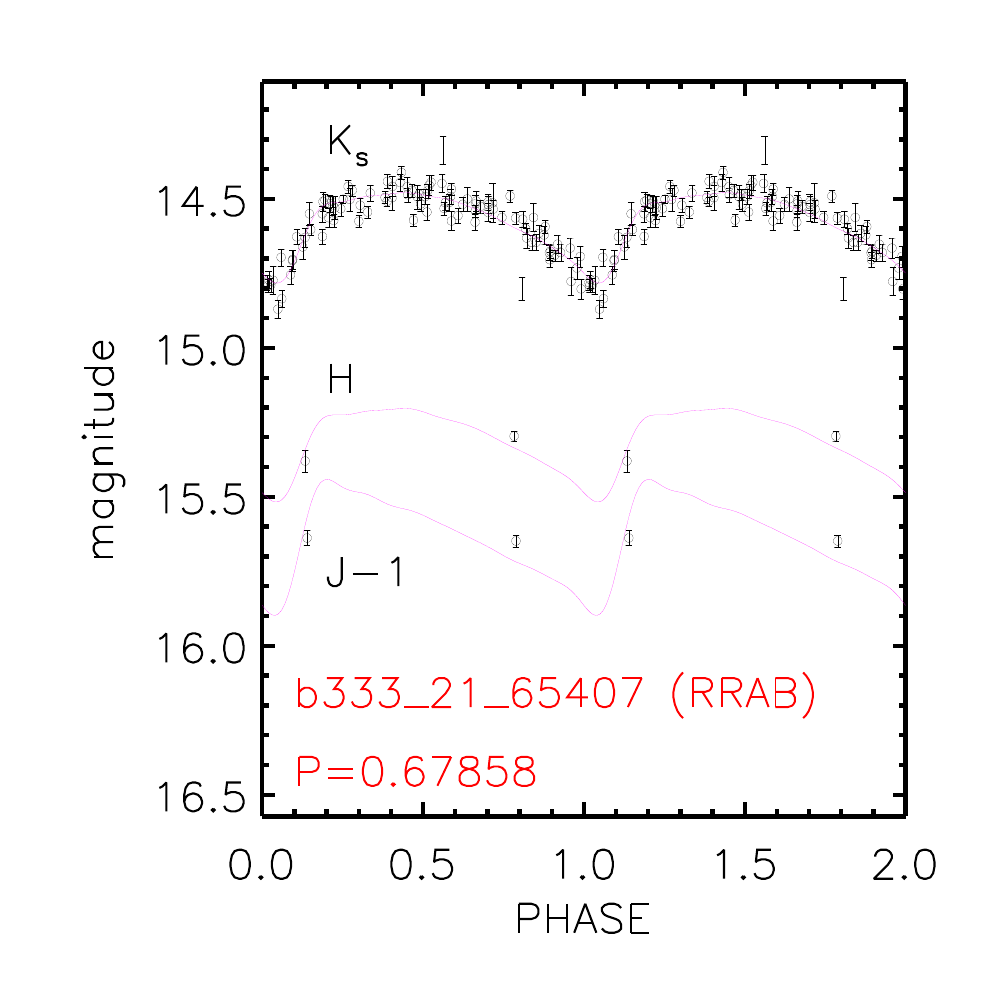}
\includegraphics[trim={1.5cm 0 0 0},clip,height=6.5cm]{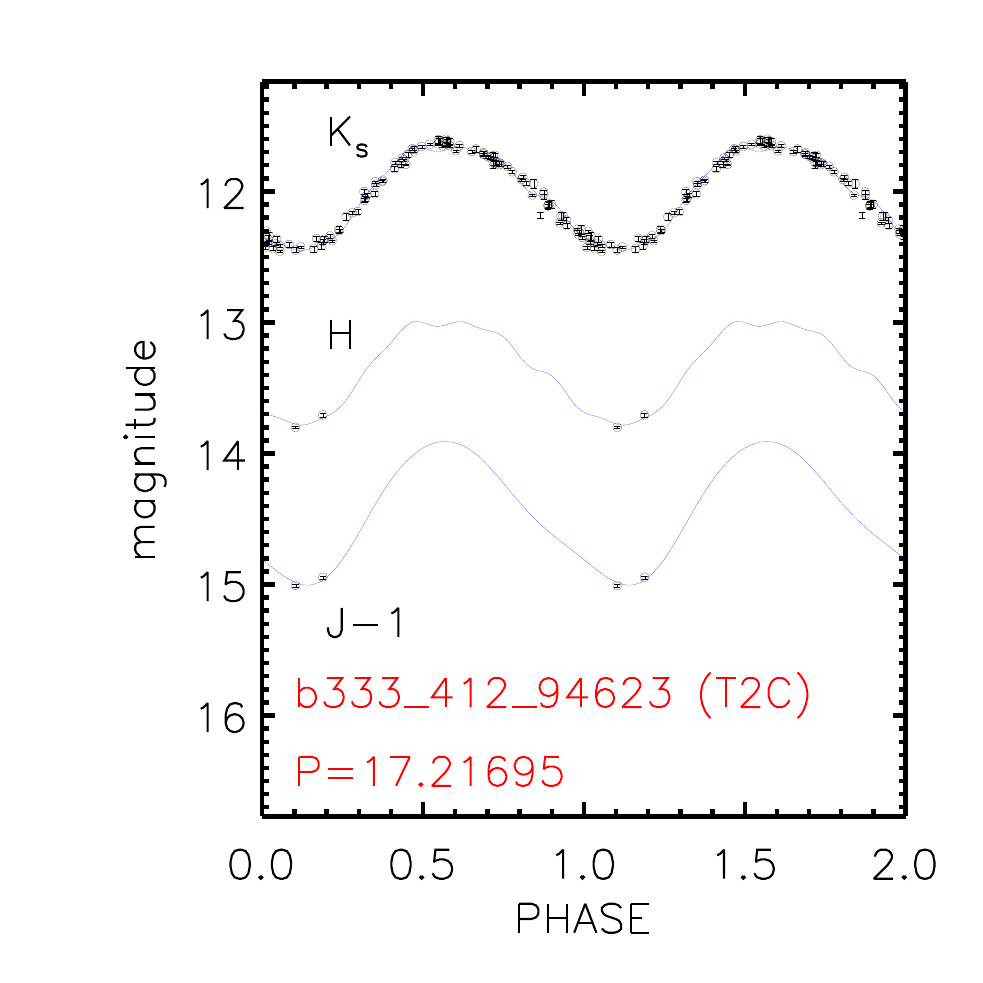}
\includegraphics[trim={1.5cm 0 0 0},clip,height=6.5cm]{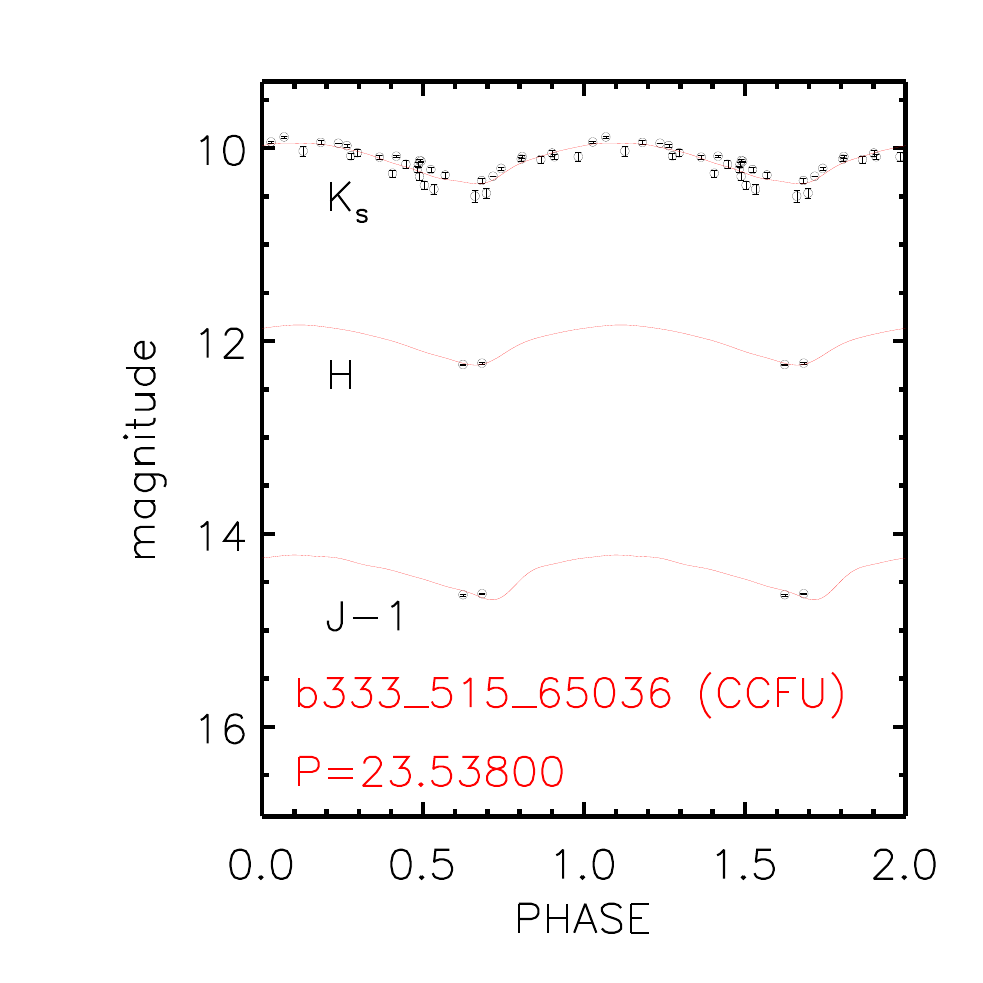}
\caption{ Left: Phased $J$-, $H$-, and $K_s$-band light curves 
of a RRab over which we applied the light-curve template.
The ID, type, and period of the star are labeled in red.
Center: As in left, but for a T2C.  
Right: As in left, but for a CCFU.}
\label{fig:template}
\end{figure*}

To perform the fits, one 
must rescale the amplitude using the $Amp(J)/Amp(K_s)$ and 
$Amp(H)/Amp(K_s)$ amplitude ratios. For RRLs and CCs, we adopted
the ratios provided by \citet{braga2018} and \citet{inno15}, respectively.
For T2Cs, we checked that both $Amp(J)/Amp(K_s)$ and $Amp(H)/Amp(K_s)$ are equal to one, within the 
uncertainties, using globular cluster data from \citet{matsunaga06}.
Furthermore, we verified that minor phase shifts between the 
$J$, $H,$ and $K_s$ light curves are negligible ($<$0.05 pulsation cycles).
Therefore, the $J$- and $H$-band template fits were performed by 
leaving only the mean magnitude as a free parameter.
Figure~\ref{fig:template} shows the use of the template 
for an RRab (left), T2C (center), and CC (right). 
Computing $\langle J \rangle$ and $\langle H \rangle$
based on the intensity integral over the template provides 
mean magnitudes that are more accurate than the simple average of the 
two epochs, and this is even more critical when the $J$- or $H$-band 
data are sampled around the minimum.

We point out that due to the high extinction, not all of 
the candidate variables have $JH$-band data with acceptable
photometric errors ($<$0.2 mag). This means that we could not 
apply the PL$HK_s$ method to all of them. More precisely, 
we lack precise $J$-band data for 16 T2Cs and 
$H$-band data for 8 T2Cs and 1 CCFU.

To sum up,  for 2 CCs, 5 RRabs, and 156 T2Cs, we derived
two sets of distances and 
extinctions which we labeled $d_{S14}$, $A_{Ks(S14)}$, and 
$d_{PL}$, $A_{Ks(PL)}$, respectively, for those derived using the 
S14 reddening map and those obtained with the PL$HK_s$ method.

% non solo non c'è il template ma nemmeno la PLH... no!!
% We adopted this method also for the ACFU candidate,
% using the fit to the $K_s$-band light curve as a pseudo-template
% and $Amp(H)/Amp(K_s)$=1. The uncertainty associated to the use 
% of the template is $\sigma_{HK} \cdot Amp(K_s)$, where $\sigma_{HK}$
% is the standard deviation of the $Amp(H)/Amp(K_s)$ ratio. We did not 
% use this technique for Miras since their larger amplitudes might 
% lead to larger errors (up to $\sim$0.1 mag).

{\it 4) Velocity} --- We use the provisional distances found
before to derive provisional transverse velocities, using 
$v_t = 4.74\cdot d \cdot \mu$, where $d$ is in kiloparsecs, 
$\mu$ in mas/yr and $v_t$ in km/s. We note that to 
derive the direction and intensity of the tangential 
velocity, we summed the 
proper motion of Sgr A* \citep{reid2004} and 
our own proper motions. According to 
\citet{braga2018b} we assume that this operation 
provides a proxy of absolute proper motions. We 
did not correct for the peculiar motion of the Sun
because we are interested in understanding whether 
our targets follow the rotation curve of the 
Galactic disk, both on our side and on the far side of it.
We assume, as a threshold
on the reliability of $\mu$, a combined statistical error 
($CSE=\sqrt{err\mu_{l*}^2+err\mu_{b}^2}$) of 2 mas/yr, 
as did \citet{contreras2017}, who used the same data.
We also use the direction of $v_t$
as a diagnostic to discriminate 
between types of Cepheids belonging 
to different Galactic populations 
(e.g., we assume that the motion of 
CCs and young variables must be disk-like). We did not include radial 
velocities ($v_r$) in our analysis for two reasons. First, 
$v_r$ from the APOGEE survey \citep{majewski2017} 
are available for only seven of our targets.
Second, and most important, with the exception of EBs and NPVs, 
our targets are radially pulsating stars, with radial-velocity
amplitudes of the order of tens of kilometres per second \citep{bono00a,feast08} 
due to the inflation and deflation of the layers where absorption 
lines originate. Therefore, a single $v_r$ measurement is not an accurate 
estimate of the radial velocity of the barycenter of the star.

{\it 5) Dereddened $H-K_s$ color} --- Generally, our 
dereddened colors have large uncertainties (the peak of the 
distribution of $(H-K_s)_0$ is at $\sim$0.3 mag), which also increase
after the dereddening. Moreover, while we have accurate 
estimates of $\langle K_s \rangle$, this is not true for 
other bands. Finally, since the reddening is extremely severe, 
we can only rely on the $H-K_s$ color, which covers a very
small wavelength range. Nonetheless, $(H-K_s)_0$ 
comes in handy to separate between
Miras and long-period CCs, since the former are $\gtsim$0.5 mag 
redder \citep{matsunaga2009b,matsunaga2013}.

{\it 6) Shape of the light curve} --- Although in the NIR many features
of the light curves present in the optical are smoothed,
it is still possible, by visually inspecting the light curves,
to refine the classification of, for example, detached EBs (DEBs), which have
narrow, alternating deep and shallow minima between almost flat plateaus.

\begin{sidewaystable*}[!htbp]
 \scriptsize
 \caption{Properties of the candidate PVS.}
  \centering
 \begin{tabular}{l@{\hspace{0.5mm}}l@{\hspace{0.5mm}}l@{\hspace{0.5mm}}c@{\hspace{0.5mm}}c@{\hspace{0.5mm}}r@{\hspace{0.5mm}}c@{\hspace{0.5mm}}c@{\hspace{0.5mm}}c@{\hspace{0.5mm}}c@{\hspace{0.5mm}}c@{\hspace{0.5mm}}c@{\hspace{0.5mm}}c@{\hspace{0.5mm}}c@{\hspace{0.5mm}}c@{\hspace{0.5mm}}c@{\hspace{0.5mm}}c@{\hspace{0.5mm}}c@{\hspace{0.5mm}}c@{\hspace{0.5mm}}c@{\hspace{0.5mm}}c}
%  \begin{tabular}{lllccrccccccccccccccc}
 \hline\hline  
ID & sat\tablefootmark{a} & Literature ID \tablefootmark{b} & $RA$ & $DEC$ & type\tablefootmark{c} & $P_{best}$ & $Z$\tablefootmark{d} & $Y$\tablefootmark{d} & $J$\tablefootmark{e} & T$_{J}$ & $H$\tablefootmark{e} & T$_{H}$ & $\langle K_s \rangle$ & $Amp(K_s)$ & $\mu_{l*}$ & $\mu_{b}$ & $A_{Ks}$ & $d$ & $v_t$\\
  & & & deg & deg &  & days & mag & mag & mag & & mag & & mag & mag & mas/yr & mas/yr & mag & kpc & km/s \\
 \hline 
 & & & & & & & & & & & & & & & & & \\
b333\textunderscore59\textunderscore98743        &    &                                          &  265.765256 & --28.941707 &     96 &    0.387960 &  20.132$\pm$0.114 &  18.622$\pm$0.032 &  16.873$\pm$0.023 &   &  15.547$\pm$0.030 &   &  14.770$\pm$0.047 &   0.252$\pm$0.049 &     0.31$\pm$1.87 &     --5.12$\pm$1.65 &    \ldots &       \ldots     &        \ldots    \\
b333\textunderscore46\textunderscore54772        &    &                                          &  265.725409 & --29.200448 &      9 &    0.406797 &  20.797$\pm$0.117 &  18.752$\pm$0.059 &  16.909$\pm$0.014 &   &  15.331$\pm$0.015 &   &  14.403$\pm$0.027 &   0.190$\pm$0.026 &   --2.45$\pm$0.55 &     --0.70$\pm$0.80 &    \ldots &       \ldots     &        \ldots    \\
b333\textunderscore16\textunderscore24211        &    &                                          &  265.689600 & --29.489227 &    469 &    0.442667 &  17.950$\pm$0.018 &  16.163$\pm$0.011 &  14.622$\pm$0.008 &   &  13.378$\pm$0.015 &   &  12.567$\pm$0.018 &   0.141$\pm$0.018 &   --2.10$\pm$0.42 &     --1.11$\pm$0.36 &    \ldots &       \ldots     &        \ldots    \\
b333\textunderscore29\textunderscore10464        &    & OGLE-BLG-ECL-069715                      &  265.506766 & --29.000586 &      A &    0.505139 &  14.058$\pm$0.005 &  13.593$\pm$0.008 &  13.126$\pm$0.007 &   &  12.805$\pm$0.010 &   &  12.482$\pm$0.017 &   0.148$\pm$0.016 &   --4.21$\pm$0.40 &     --1.98$\pm$0.36 &    \ldots &       \ldots     &        \ldots    \\
b333\textunderscore38\textunderscore13973        &    &                                          &  266.087807 & --29.593169 &    469 &    0.505688 &         \ldots    &  20.514$\pm$0.159 &  18.547$\pm$0.059 &   &  16.530$\pm$0.035 &   &  15.185$\pm$0.041 &   0.260$\pm$0.043 &   --6.64$\pm$1.59 &     --1.12$\pm$1.50 &    \ldots &       \ldots     &        \ldots    \\
b333\textunderscore414\textunderscore55144       &    & [MCZ2016]\textunderscore VVV-RRL-55144   &  266.104489 & --28.659180 &      5 &    0.508390 &         \ldots    &         \ldots    &  17.936$\pm$0.083 & * &  15.829$\pm$0.054 & * &  14.635$\pm$0.043 &   0.406$\pm$0.047 &  --11.57$\pm$1.63 &       2.67$\pm$1.88 &      1.33 &    5.34$\pm$0.24 &   300$\pm$32     \\
b333\textunderscore511\textunderscore117170      &    &                                          &  266.345607 & --29.183243 &     95 &    0.515417 &         \ldots    &  20.243$\pm$0.294 &         \ldots    &   &  17.946$\pm$0.177 &   &  16.031$\pm$0.128 &   0.257$\pm$0.128 &   --8.92$\pm$5.74 &     --3.56$\pm$3.80 &    \ldots &       \ldots     &        \ldots    \\
b333\textunderscore25\textunderscore81044        &    &                                          &  265.468647 & --29.324717 &     59 &    0.517853 &         \ldots    &  21.041$\pm$0.442 &  19.758$\pm$0.129 &   &  17.007$\pm$0.055 &   &  15.505$\pm$0.065 &   0.274$\pm$0.065 &   --1.69$\pm$2.18 &     --1.94$\pm$1.98 &    \ldots &       \ldots     &        \ldots    \\
b333\textunderscore53\textunderscore12695        &    &                                          &  265.773550 & --29.685386 &     A5 &    0.524798 &         \ldots    &         \ldots    &  18.228$\pm$0.049 &   &  16.372$\pm$0.034 &   &  15.265$\pm$0.048 &   0.306$\pm$0.051 &   --3.17$\pm$1.74 &       0.52$\pm$1.67 &    \ldots &       \ldots     &        \ldots    \\
b333\textunderscore44\textunderscore75303        &    &                                          &  266.006268 & --29.936152 &     94 &    0.525078 &  19.394$\pm$0.047 &  17.550$\pm$0.016 &  15.904$\pm$0.011 &   &  14.403$\pm$0.014 &   &  13.278$\pm$0.028 &   0.190$\pm$0.027 &   --2.63$\pm$0.55 &       0.21$\pm$1.28 &    \ldots &       \ldots     &        \ldots    \\
\hline
 \end{tabular}
 \tablefoot{Only the first ten rows of the table are displayed. The full table is
 provided in the machine-readable version of the paper.
 \tablefoottext{a}{An asterisk indicates that the variable is 
 most likely heavily saturated ($K_s<$11 mag), therefore its magnitudes 
 and amplitude might be inaccurate.}
 \tablefoottext{b}{ Alternative IDs of the variables from
 the literature (see Section~\ref{sect_vsxogle}).}
 \tablefoottext{c}{0: CCFU; 1: CCFO; 2: T2C; 3: ACFU; 4: ACFO; 
 5: RRab; 6: RRc; 7: SRV; 8: Mira; 9: NPV; A: DEB. 
 A ``--'' sign indicates a low-rank variable, meaning that its light 
 curve is noisy and the classification is not certain.}
 \tablefoottext{d}{The listed $ZY$ magnitudes are obtained from
 the average of the two available phase points.}
 \tablefoottext{e}{ The listed $J$ and $H$ magnitudes are obtained either from
 the average of the two available phase points or applying the template. 
 In the latter case, the columns ``T$_{J}$'' and/or 
 ``T$_{H}$'' are marked with an ``x''.
 The photometric and pulsation properties of Miras---except those
 in common with M09---are not within this table and will be 
 published in Nikzat et al. (2019, in prep.)}.}
\label{tab:dist}
\end{sidewaystable*}

We provide the light curves of these 1,019 candidate variables 
in Table~\ref{tablcv}.

% per tirare giù le cdl
% awk '{print "ls "$1"/"$1"_"$2"_temp3.fas"}' lista_cepheids_good_temp3_v10_refined3 > schifo
% sostituire ls con awk '{print $1,"&",$3,"&",$4}' in schifo
% ora uso temp_190214 per capire quando si può usare la distanza e quale usare

\begin{table*}[!htbp]
 \footnotesize
 \caption{$K_s$-band time series of our candidate variables.}
  \centering
 \begin{tabular}{l c c c}
 \hline\hline  
Name &  HJD--2,400,000 & Mag & Err \\ 
 & days & mag & mag \\ 
\hline 
  b333\textunderscore59\textunderscore98743 & 55844.02155 & 14.704 & 0.043 \\
  b333\textunderscore59\textunderscore98743 & 55423.14904 & 14.607 & 0.040 \\
  b333\textunderscore59\textunderscore98743 & 55777.13660 & 14.865 & 0.054 \\
  b333\textunderscore59\textunderscore98743 & 55778.19837 & 14.627 & 0.032 \\
  b333\textunderscore59\textunderscore98743 & 55794.12621 & 14.657 & 0.047 \\
  b333\textunderscore59\textunderscore98743 & 55806.15221 & 14.674 & 0.046 \\
  b333\textunderscore59\textunderscore98743 & 55820.10395 & 14.705 & 0.045 \\
  b333\textunderscore59\textunderscore98743 & 55830.02354 & 14.807 & 0.044 \\
  b333\textunderscore59\textunderscore98743 & 55849.02315 & 14.518 & 0.036 \\
  b333\textunderscore59\textunderscore98743 & 55987.37349 & 14.730 & 0.052 \\
  \hline
 \end{tabular}
 \tablefoot{Column 1 gives the name, 
 column 2 the heliocentric Julian day of the observation, column 3 the measured magnitude and 
column 4 the photometric error. Only the first ten entries are 
listed. The full table is provided in electronic form. We note that the 
light curves of Miras---except those in common with M09---are not 
within this table and will be 
 published in Nikzat et al. (2019, in prep.)}
\label{tablcv}
\end{table*}

% ;variabili bona-fide
% ;awk '{print $4}' lista_cepheids_good_temp3_v10 | grep -v '-' | awk 'length($1)==1 {print $0}' |wc -l
% ;variabili senza "?" ma ancora incerte
% ;awk '{print $4}' lista_cepheids_good_temp3_v10 | grep -v '-' | awk 'length($1)!=1 {print $0}' |wc -l

After taking account of the quoted criteria, we assigned a
variable type to each of the candidate PVSs in the final sample.
We ended up with unambiguous classification for 472 variables (5 RRab, 
164 T2Cs, 3 CCFU, 1 ACFU, 16 SRVs, 210 Miras and 73 NPVs, 
of which 47 DEBs, see Table~\ref{tab:dist},
second column). For the other 547 variables, we could not provide a
solid classification due to uncertainties on the reddening and distance, 
and due to noisy light curves. More specifically, for 99 of them, 
the light curve is clear enough to be classified as ``high-rank''
candidate variables. However, due to the lack of other information, 
or ambiguity in the criteria listed above, 
we could only provide a tentative classification (e.g., 96 in the
fifth column in Table~\ref{tab:dist} means that we are uncertain
on whether b333\textunderscore509\textunderscore 98743 
is a RRc or a NPV). For the remaining 448 candidate PVSs, 
the light curve was not clear enough (due to noise, high 
photometric error, or secondary modulations), and therefore they 
were classified as ``low-rank variables''. These too have to 
be considered as variables without a certain classification.

We point out that we have found the first bona fide AC towards 
the Galactic Bulge at such low latitudes. Its position in the Bailey diagram and 
the shape of its light curve were crucial to providing a
solid classification. We have submitted a 
follow-up proposal to collect spectroscopic data with 
FIRE@Magellan, to estimate radial velocity and iron
abundance for this star. Anomalous Cepheids are relatively rare objects, especially in the 
Milky Way, and spectroscopic information is available
only for one of them \citep[V19 in the globular cluster
NGC 5466,][]{mcarthy97}.

\begin{figure*}[!htbp]
\centering
\includegraphics[width=24cm,angle=90]{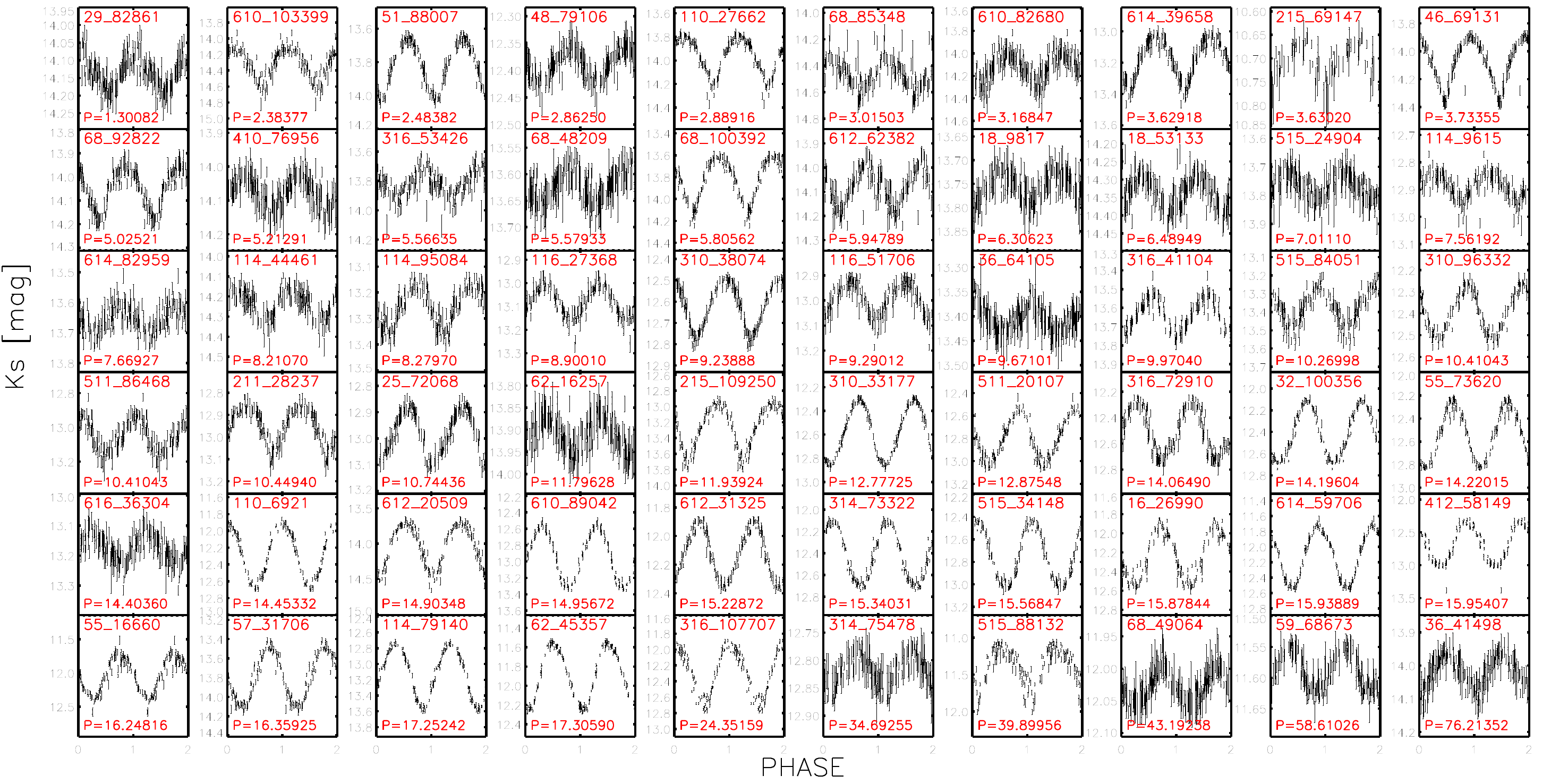}
\caption{Phased $K_s$-band light curves of 60 variable stars classified as T2Cs. 
Periods are labeled at the bottom of each panel. The ``b333'' part 
of the name is omitted.}
\label{fig:allcvs_t2c}
\end{figure*}

\begin{figure*}[!htbp]
\centering
\includegraphics[width=19cm]{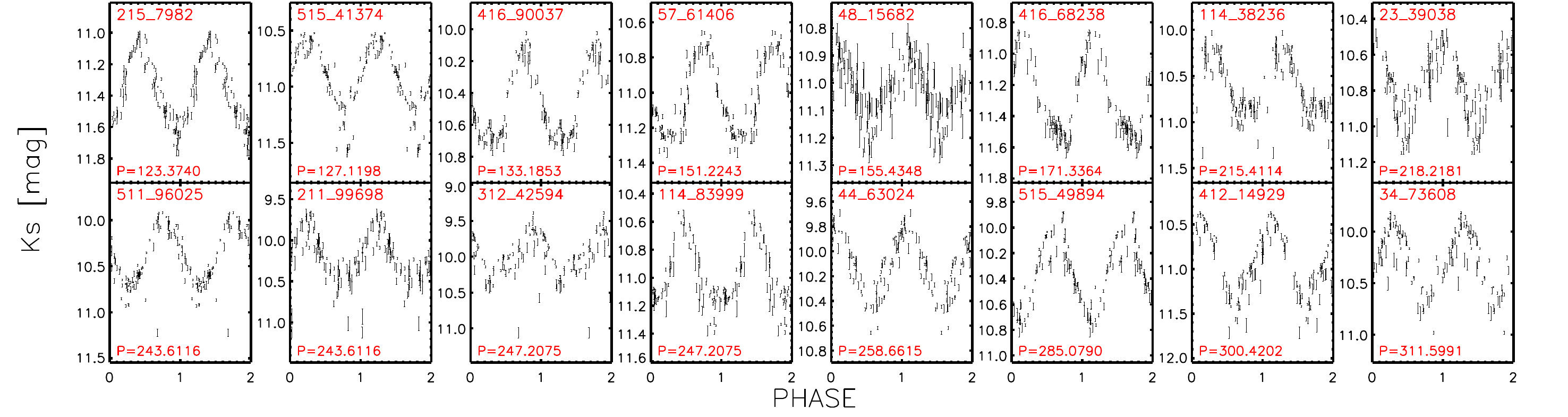}
\caption{ As in Fig.~\ref{fig:allcvs_t2c} but 
for the 16 variable stars in common with the sample of M09. 
We note that, while all of these were classified as Miras by M09, 
our classification does not always match. For a more detailed
discussion,  see Section~\ref{sect_matsu09}.}
\label{fig:allcvs_mira}
\end{figure*}

Figure~\ref{fig:allcvs_t2c} displays a sample of light curves 
of the bona fide T2Cs that we have detected. Figure~\ref{fig:allcvs_mira}
is the same for Miras and SRVs in common with M09.

% awk '{printf "%20s  %10.6f  %10.6f  %5s  %12.6f  %7.3f %7.3f  %7.3f %7.3f  %7.3f %7.3f  %7.3f %7.3f  %7.3f %7.3f  %7.3f %7.3f  %9.3f  %9.3f %9.3f  %9.3f  %9.3f  %9.3f  %9.3f  %9.3f  %9.3f  %9.3f  %9.3f  %9.3f  %9.3f  %9.3f  %9.3f  %9.3f  %9.3f  %9.3f  %9.3f  %9.3f  %9.3f  %9.3f  %9.3f  %9.3f  %9.3f  %9.3f  %9.3f  %9.3f  %9.3f  %9.3f  %9.3f  %9.3f \n", "b333_"$1"_"$2,$145,$146,$3,$7,$8,$19,$9,$20,$10,$21,$11,$22,$13,$24,$39,$40,$158,$159,$160,$161,$162,$163,$164,($158^2+$160^2)^.5,$41,$42,$46,$54,$55,$59,$67,$68,$72,$80,$81,$85,$93,$94,$98,$106,$107,$111,$119,$120,$124,$132,$133,$137}' lista_cepheids_good_temp3_v10_refined2 > schifo2
% #     b333_#paw_ID       ra           dec       type         per     Z        eZ        Y        eY      J      eJ        H      eH     Kmean    eKmean   AmplK   eAmplK    pm_lcb   errpm_lcb     pm_b      errpm_b  ak_mats 0  dmatsu  errdmatsu      dist_rrc  edist_rrc   ak_rrc    dist_rrab edist_rrab    ak_rrab dist_acfo edist_acfo    ak_acfo dist_acfu edist_acfu     ak_acfu dist_ccfu edist_ccfu    ak_ccfu    dist_ccfo edist_ccfo   ak_ccfo   dist_t2c edist_t2c     ak_t2c    dist_mira edist_mira   ak_mira
% ora uso temp_190214 per capire quando si può usare la distanza e quale usare

% per le figure: IDLWorkspace/Default/Temp/temp_180911.pro

\section{Matches with other catalogs}

Although this work presents the most extensive survey of this region of the 
Galaxy for variables with periods longer than one day, and despite 
most of our targets being new detections, some were already found in similar
investigations. In the following paragraphs, we compare our list with those found 
previously in the literature, more specifically, within the VSX 
catalog, the OGLE survey and---in order of increasing variable period 
(from RRLs, T2Cs, CCs to Miras)---CR18, M13, and M09. We did not find
any match with the variables found by \citet{dong2017} within the 
nuclear star cluster.

\subsection{VSX and OGLE catalogs}\label{sect_vsxogle}

We crossmatched our list of candidate variables with those in the 
VSX catalog and those found by the OGLE survey to see whether or not some of our 
variable candidates had already been found and/or classified. 
Indeed, we found nine matches within the eclipsing binaries catalog of 
the OGLE IV survey \citep{soszynski2016b} and 77 matches in the 
VSX catalog. The literature names of these matching variables are 
displayed in the second column of Table~\ref{tab:dist}.

The crossmatches were performed by selecting a conservative
radius for the cone search of 3 arcsec. The VSX matches were all 
found within 2.4 arcsec, while the OGLE matches are within 1.4 
arcsec from our own position. We checked that the quoted matches
are indeed the same objects that were found in our investigation by
comparing not only the distance from our position, but also the 
magnitude and, when needed, the finding charts. The matching 
sources in the VSX catalog were originally detected by \citet{wood1998}, \citet{glass2001};
M09, \citet{matsunaga11} and M13. More precisely, we find 15 sources in common with 
\citet{wood1998}, 2 in common with \citet{glass2001} , 
and the remaining 60 are the CCs by \citet{matsunaga11}, the 16
T2Cs by M13, 7 eclipsing/unknown type variables by M13, and 34 LPVs
by M09. Curiously, only 4 bona fide Miras by M09 were retrieved 
in the VSX catalog, although we found 16 common variables by directly
matching our catalog to the catalog of bona fide Miras of M09 (see
Section~\ref{sect_matsu09}). By summing up all this information, 
we conclude that the variables that were already
known before this work are 77 VSXs,  9 OGLE EBs,  12 
M09 Miras (not retrieved within the VSX), and 3 
RRLs  \citep[][see Section~\ref{sect_rrl_rodrigo}]{contreras2018},
making a total of 101.

% Match vsx: vsx_match.csv in ~/Documenti/Surveys/VVV/data/tile/b333/phased_lcv/
% temp_190311.pro

We conclude this section by pointing out that our independent 
classification accurately matches those in the literature. 
In fact, the candidate variables matching the 
catalogs of \citet{wood1998} and \citet{glass2001} were all classified
as either OH stars or Miras; our classification for these 
stars is either ``Mira'' or ``Mira?'', thus matching the classification 
provided by the quoted studies. Concerning the matches with variables
in the M09, \citet{matsunaga11}, and M13 series of studies, more details 
are provided in Sects.~\ref{sect_matsu13} and ~\ref{sect_matsu09}.

\subsection{RR Lyr catalog of CR18}\label{sect_rrl_rodrigo}

As discussed in Section~\ref{sect_freqgrid}, our periodicity search 
algorithm focused on periods longer than 1 day, and the variables
with shorter periods are an incomplete sample. Nonetheless, our final catalog
contains some candidate RRLs that are either uncertain or bona fide. We matched our catalog
with the list of RRLs published by CR18 and found three sources in common. We note that 
we performed the match by ID because we used the same photometric
data set. This means that there is no uncertainty associated to a cone 
search by coordinates.

The common sources are b333\textunderscore616\textunderscore55278, 
b333\textunderscore614\textunderscore37068 and
b333\textunderscore414\textunderscore55144, which we classified as RRab/NPV/ACFU, RRab/ACFU, and RRab,
respectively. Also, we 
note that in our catalog, we found four RRabs that
were not detected by CR18 
(b333\textunderscore201\textunderscore84779, 
b333\textunderscore304\textunderscore81788, 
b333\textunderscore201\textunderscore45267 and 
b333\textunderscore201\textunderscore65407).

% temp_180928 per cercare i match

\subsection{T2C and CC catalog of M13}\label{sect_matsu13}

A previous search for Cepheids towards the Galactic center was 
performed by M13 with the NIR SIRIUS camera at the IRSF 1.4m telescope.
They found 45 variable stars, of which 20 were classified as Cepheids
(16 T2Cs, 3 CCs and one generic Cepheid candidate).

The b333 tile overlaps completely with the IRSF survey sky area and we
retrieved all the 20 Cepheids by M13 within our initial 
list of $\sim$5 million sources. However, our periodicity
search algorithm detected only 14 of them. Six Cepheids were not
retrieved because their light curves have many uncertain
phase points due to either saturation and/or blending.
We manually extracted the light curves
of the six missing Cepheids and, by knowing a priori their 
periods, we manually rejected the poor-quality data and obtained 
clean light curves for five of them. Unfortunately,
b333\textunderscore114\textunderscore65025 is affected
by severe blending and it was not possible to detect any periodic 
behavior even knowing the period a priori. 
We note that M13 also  pointed out that this star 
(their \#2) is in a very crowded field and that their photometry was 
not accurate.

The comparison with the M13 sample also provides a validation of our 
classification criteria. In fact, among the 14 variables that were retrieved
automatically, we  classified -- without knowing, a priori, the classification 
by M13 -- 13 of them as T2Cs and one as ``T2C?''. We  then 
checked the two samples and found that our classifications match
those of M13, including their ``Cep(?)'', which is our ``T2C?''.

We provide a more detailed discussion of the offsets in distance and reddening between 
our estimates and those of M13 in Section~\ref{sect_caveat}.

\subsection{Miras catalogs of M09}\label{sect_matsu09}

Using the NIR SIRIUS camera at the IRSF 1.4m telescope, M09 found 
175 Miras towards the GC, and estimated their 
periods, distance, and extinction. They performed a sample 
selection based on the period, and marked only
those with periods between 100 and 350 days as bona fide Miras, since they are the 
least affected by circumstellar reddening. We found a matching source
for all their Miras in our initial photometric catalog. However, only 
16 of them were retrieved in our final catalog of variable candidates.
The reason for the low fraction of Miras retrieved in our 
final catalog is that M13 Miras are mostly located in the Bulge, 
therefore they are saturated -- more severely than T2Cs -- or 
are in the nonlinear regime of the VISTA images. A more complete 
discussion of these and other Mira variables in the VVV data will
be presented in Nikzat et al. (2019, in preparation).

Of these 16 variable candidates, 9 were classified as Miras, 
6 as ``Mira?'', and one as an SRV. We point out that the offset
in mean magnitude between our catalog and that of M13 is small:
the mean offset is 0.002$\pm$0.092 mag, although for one 
variable (b333\textunderscore416\textunderscore5350, M09 \#1043) 
the offset is as large as 0.218
magnitudes brighter in M13. However, the offsets on $Amp(K_s)$
are larger. As a matter of fact, for nine among the 16 variables, 
the offset in $Amp(K_s)$ is larger than 0.1 mag, and for two of them 
it is as large as $\Delta A_{Ks} \sim$2.0 mag, which has dramatic 
effects on the distance estimates. We found that $\Delta Amp(K_s)$ 
is not correlated with the mean magnitude.

% {\bf Nine Miras were also detected by NM18 using VVV data 
% in the region of the Arches and Quintuplet stellar clusters... 
% waiting RA, DEC from Navarro Molina for comparison.}

We provide a more detailed discussion of the offsets in distance and reddening between 
our estimates and those of M09 in Section~\ref{sect_caveat}.

\subsection{A caveat on extinction and distance}\label{sect_caveat}

\begin{figure*}[!htbp]
\centering
\includegraphics[width=12cm]{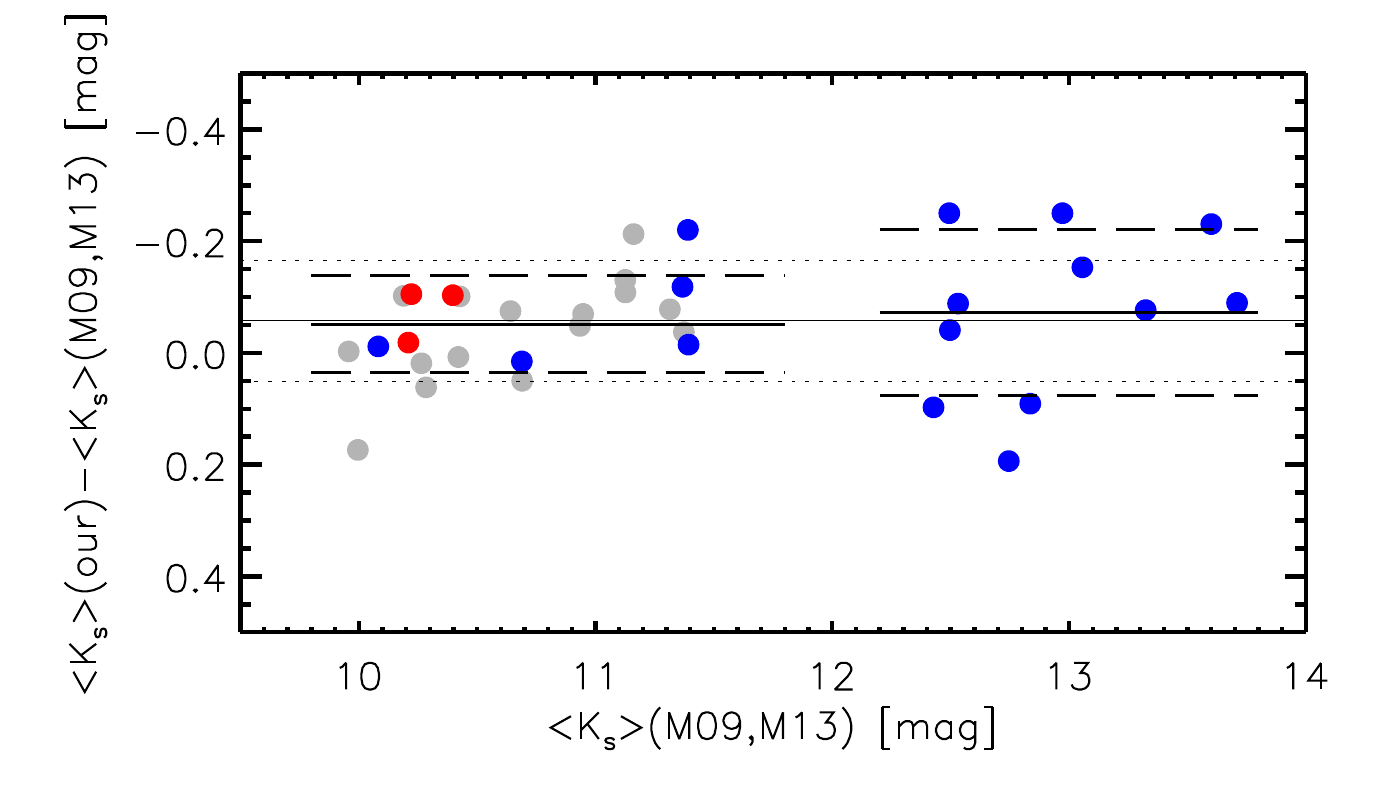}
\caption{Mean magnitude offset between our magnitudes and those
by M09 and M13 ($\langle K_s \rangle (M09,M13)$). 
Blue circles display T2Cs, red circles
display CCs, and gray circles display Miras.
The black solid line represents the average, and the 
black short-dashed lines the standard deviation of the whole sample.
The black, thick, solid, and long-dashed lines have the same meaning for
saturated ($\langle K_s \rangle <$12 mag) and 
nonsaturated ($\langle K_s \rangle >$12 mag) targets in the VVV.}
\label{fig:saturation}
\end{figure*}

In principle, the offsets in distance are not to be fully ascribed to 
offsets in reddening. As mentioned in Section~\ref{selectiontarget}, IRSF and 
VISTA have different diameters (1.4m vs 4m). This means that 
stars in the linear photon count regime of the SIRIUS camera 
on IRSF can be saturated in VIRCAM. Figure~\ref{fig:saturation}
displays the difference in mean magnitude for the common sources. 

%per fare queste figure ho bisogno del file mira_matsunaga_checktopcat... ma come lo creo?
%fare crossmatch di mira_matsunaga_checktopcat con le nuove tabelle output, prima mPL e poi S14
%salvare in mira_matsunaga_master.csv
% awk -F',' '{if ($23=="C2") print $1,$2,$3,$4,$5,$6,$7,$8,$9,$10,$11,$47,$54,$14,$15,$256,$261,$18,$19,$20,$21,$22,$23,$24,$42,$44,$27,$28,$29,$127,$122; else if ($23=="C1") print $1,$2,$3,$4,$5,$6,$7,$8,$9,$10,$11,$47,$54,$14,$15,$234,$239,$18,$19,$20,$21,$22,$23,$24,$42,$44,$27,$28,$29,$105,$100; else if ($23=="M") print $1,$2,$3,$4,$5,$6,$7,$8,$9,$10,$11,$47,$54,$14,$15,$267,$272,$18,$19,$20,$21,$22,$23,$24,$42,$44,$27,$28,$29,$138,$133}' mira_matsunaga_master.csv > mira_matsunaga_190404
%figure da IDLWorkspace/Default/Temp/temp_181024.pro

Within the dispersion, which is very large ($\sim$0.1 mag),
the average is zero. However, somewhat surprisingly,
both the offset and the dispersion are larger for nonsaturated VVV targets than
for saturated targets. The large individual offsets might be due to 
the different pixel scales of the two cameras: 
0.25 mas/pix for VIRCAM and 0.45 mas/pix for SIRIUS. In such a crowded
region, VIRCAM has the advantage of a spatial resolution that is 
almost twice as great as that of SIRIUS, meaning that it is more capable 
of resolving neighboring stars. Overall, we can 
conclude that the different saturation levels and pixel scales do not  significantly
affect the mean magnitudes, but the effect on 
individual stars is large, up to $\pm$0.2 mag, meaning a 
relative systematic offset of up to $\sim$8\% on the distance.

Once it had been found that the offsets in $K_s$ have a minor 
impact on distances, we inspected the effects of the offsets in
extinction. As mentioned in Section~\ref{sect_classification}, 
for targets like CCs and T2Cs, we  derived two sets of distance 
and extinction. We compare the $A_{Ks(PL)}$ and $d_{PL}$ with those
from M09 and M13. The main difference is that they used the 
\citet{nishiyama2006,nishiyama2009} reddening law, which 
provides a different $\dfrac{A_{Ks}}{E(J-K_s)}$ ratio from ours 
\citep[0.499 versus the 0.428 average value in ][]{alonsogarcia2017}.
This means that for a homogeneous comparison we need to 
compare $E(J-K_s)$ and not $A_{Ks}$.

\begin{figure*}[!htbp]
\centering
\includegraphics[trim={0 0 0.5cm 0},clip,height=7cm]{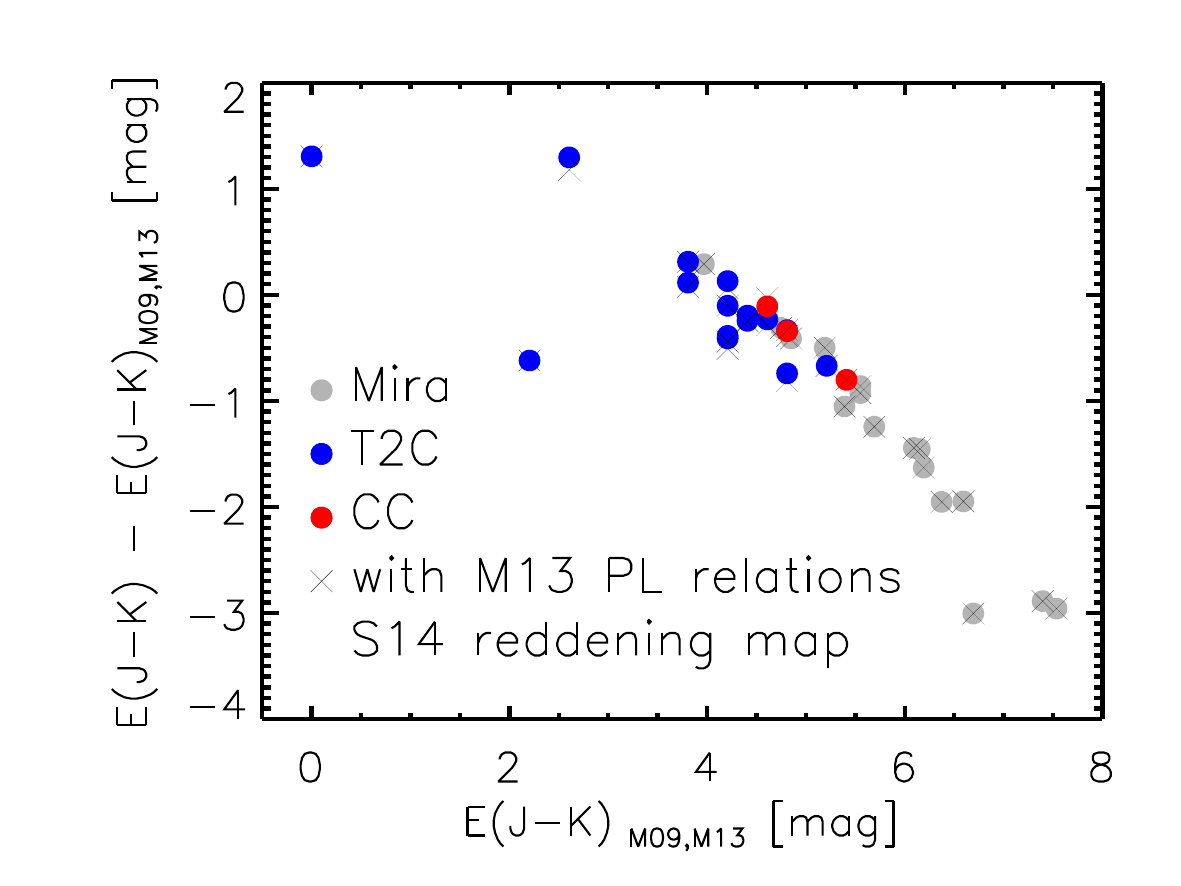} 
\includegraphics[trim={2.2cm 0 0 0},clip,height=7cm]{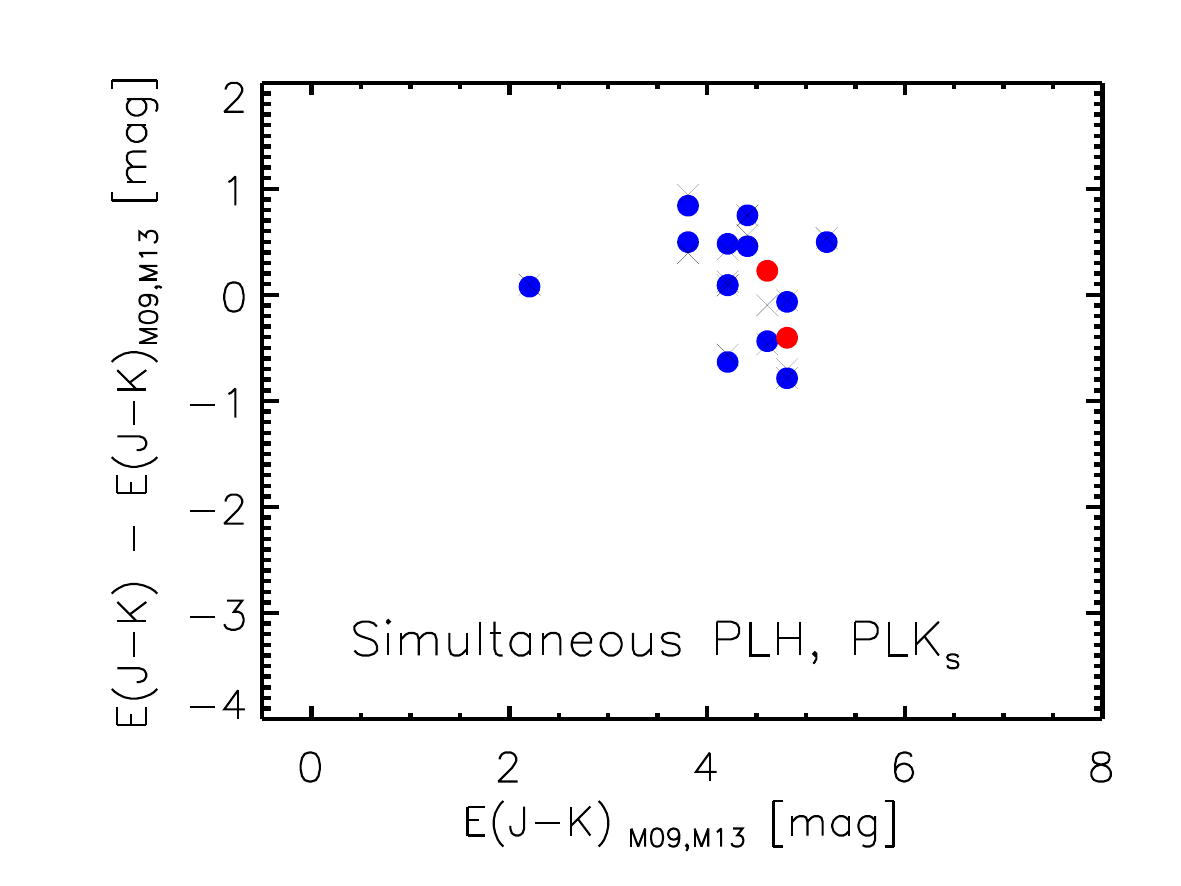}
\caption{Left: Offset between our reddening $E(J-K_s)$---obtained 
with the S14 reddening map---and that 
by M09 and M13 ($E(J-K_s)_{M09,M13}$) vs. $E(J-K_s)_{M09,M13}$
 for the common targets. The symbols
are the same as those in Fig.~\ref{fig:saturation}.
The black crosses
mark the same information, but using the same calibrating PLs
as M13. Right: As in the left panel but for our $E(J-K_s)$ derived using
the PL$HK_s$ solution.}
\label{fig:reddening}
\end{figure*}

For a rigorous interpretation of Fig.~\ref{fig:reddening} one should take into account 
that in Section~\ref{sect_classification} we have adopted
the same PL$K_s$ used by M09 for Miras, but different PL relations for 
T2Cs and CCs. Therefore, Fig.~\ref{fig:reddening} also shows our 
$E(J-K_s)$ estimates both using the PLs listed in Table~\ref{tab:vartypes}
(circles in Fig.~\ref{fig:reddening}) and using the same relations as M09 and M13,
converted to the VISTA photometric system (crosses in
Fig.~\ref{fig:reddening}). We note that to adopt the different 
calibrations of the PLs does not qualitatively change the observed 
trends. Furthermore, using the S14 map (left panel), 
our $E(J-K_s)$ estimates are always smaller
than 5 mag, while those by M09 and M13 can be as large 
as $\sim$7.5 mag and $\sim$5.5 mag, respectively. These offsets
generate the displayed trend: using the 
PL$HK_s$ solution for distance and $A_{Ks}$, 
the trend does not show up and the average offset is zero, 
within the standard deviation. Unfortunately, not all the common
targets have a photometric PSF solution in our $H$ band images, 
since they are extremely faint, and therefore the sample is smaller. 
The fraction of targets with a PSF solution in the $J$ band 
drops dramatically, and it is pointless to perform the 
same analysis on them. We remind the reader 
that we did not adopt this technique for Miras (see Section
\ref{sect_classification}, point {\it 3c}).

Most probably, the resolution of the S14 reddening map 
($6 \times 6$ arcmin) is insufficient 
to reproduce the fine details of the reddening patterns towards the 
GC, and the average values seem to be, overall, underestimated.
One might expect that adopting the G12 
reddening map, which has a resolution of 2 arcmin, would diminish this 
effect, but the average offset is negligible 
and the individual distances are even less reliable. In fact, using the G12 map, 
T2Cs display a very unlikely double-peaked distance distribution.

\begin{figure*}[!htbp]
\centering
\includegraphics[width=7.8cm]{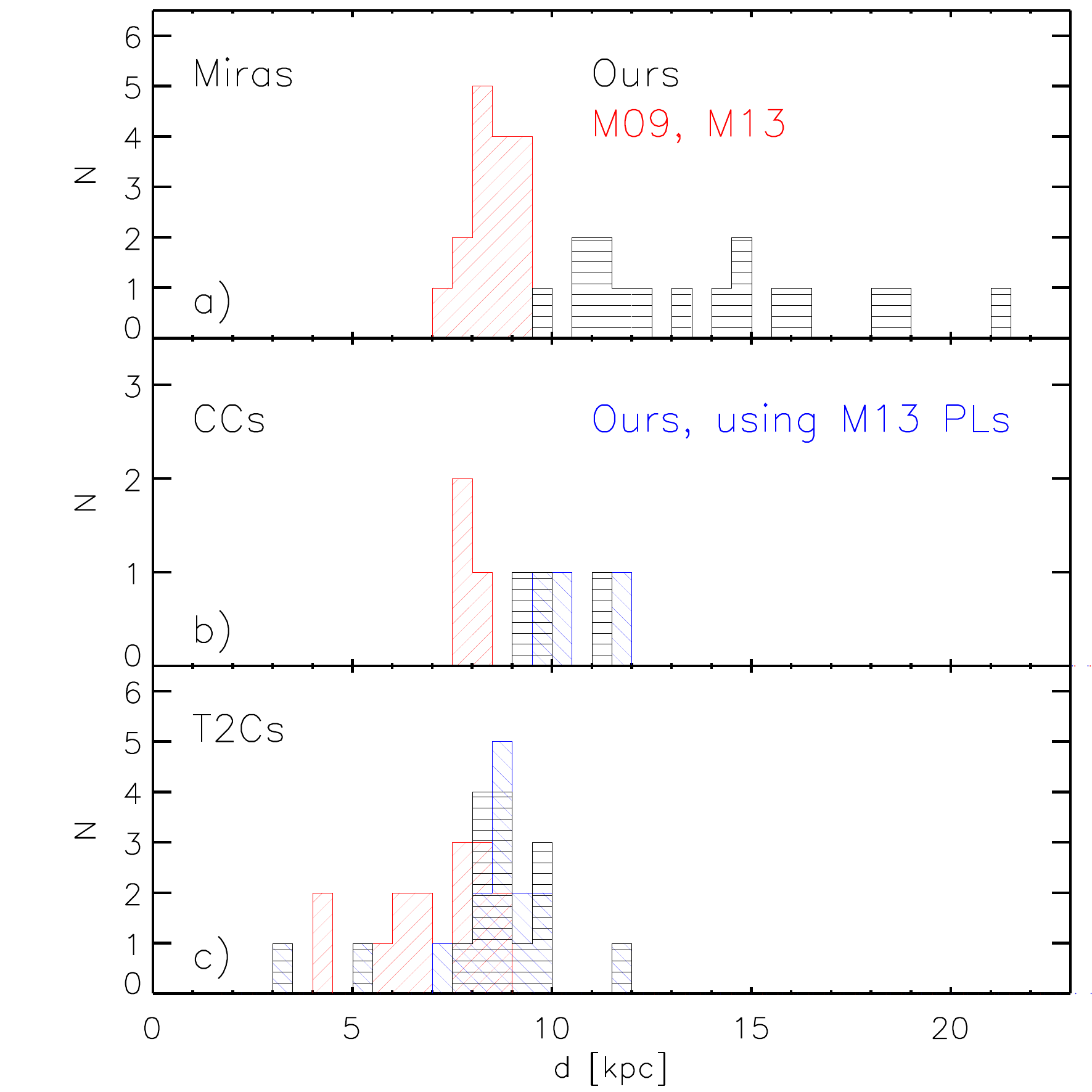}
\includegraphics[width=7.8cm]{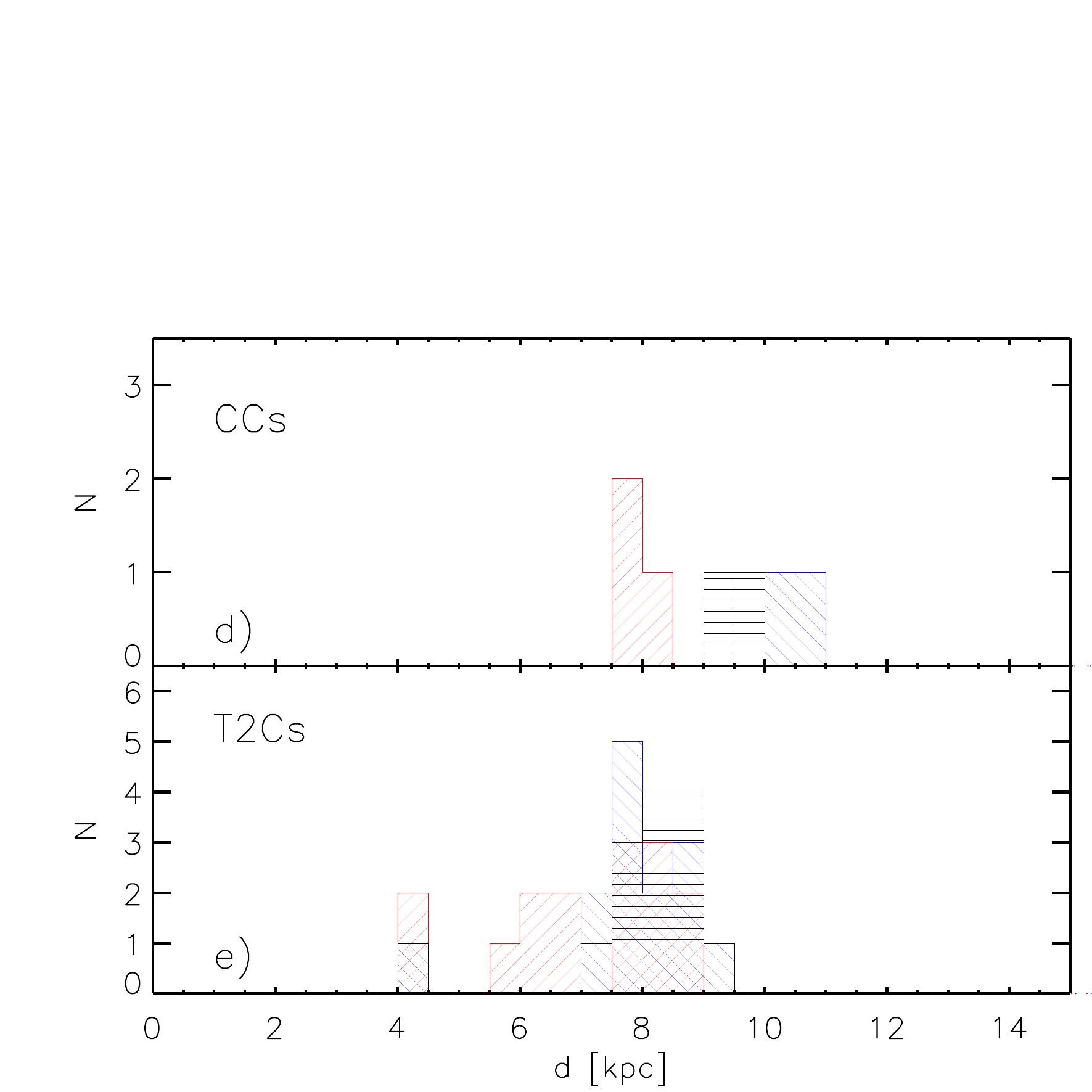}
\caption{Panel a): Distance distribution of Miras. Red: M09 distances, Black: 
our distances. Panel b): as in panel (a) but for CCs. Blue bars mark the 
distances derived adopting the same PL as M13. Panel c): as in panel (b)  
but for T2Cs. Panel d): as in panel (b) but using the distances estimated 
with the PL$HK_s$ relations. Panel e): 
as in panel (c) but using the distances estimated 
with the PL$HK_s$ relations.}
\label{fig:reddening2}
\end{figure*}

These offsets in reddening translate into significant distance offsets, 
which become dramatic for Miras, as displayed in Fig.~\ref{fig:reddening2}.

{\it Miras} --- The difference in reddening---and, in turn, on
distances---is, on average, larger for Miras than for 
T2Cs and CCs. M09 estimated the distances to 
these targets to be smaller than 10 kpc for all of them. Most of them 
are located at between 7 and 10 kpc, and their peak is around 8.5 kpc,
indicating that they belong to the Bulge. On the other hand, our 
distances are all larger than 10 kpc and their spatial distribution is almost
constant over the 10-25 kpc distance range. One might invoke circumstellar
extinction  -- which is accounted for in M09 extinction estimates, but not in the S14 
map -- to justify these offsets. However, as already mentioned, Miras with 
such short periods ($<$350 days) should not be 
significantly affected by circumstellar 
extinction \citep{itamatsunaga2011}. This means that, 
while possibly present, circumstellar extinction cannot, by itself,
account for the whole offset. As mentioned above, we did not 
adopt the PL$HK_s$ solution for Miras.

{\it CCs} --- These three Cepheids were first found by 
\citet{matsunaga11}, who claimed that they are Bulge stars, and 
the first CCs ever found close to the GC. However, our 
different extinctions provide larger distances, which put them on the 
other side of the thin disk, at more than 9 kpc, both for 
$d_{S14}$ and $d_{PL}$, with small differences ($\lesssim$0.4 kpc) between the two distance sets. 
We point out that, although these targets are saturated 
in the VVV, the difference between our $\langle K_s \rangle$
and those by M13 are small (--0.056 mag, 0.021 mag and --0.063 mag, for
b333\textunderscore215\textunderscore87770, 
b333\textunderscore215\textunderscore86376 and 
b333\textunderscore515\textunderscore65036, respectively, where negative values indicate 
brighter $\langle K_s \rangle$ in our catalog). This means that the 
distance offsets displayed in the middle panel of 
Fig.~\ref{fig:reddening2} are not caused by a poor estimate of their mean 
magnitude, but should be ascribed to the difference in extinction only.
We note that the adoption of the same PL as M13 affects the distances at a level of 
$\lesssim$0.3 kpc for the $d_{S14}$ sample  -- which is negligible compared 
to our errors -- and $\sim$1 kpc for $d_{PL}$.

{\it T2Cs} --- For all but three targets, we obtain a 
$d_{S14}$ that is larger by 1-2 kpc with respect to M13 distances, 
although a few objects show distances that are similar or smaller 
(b333\textunderscore215\textunderscore69147, which is M13 \#19; 
b333\textunderscore416\textunderscore9672, which is M13 \#34; 
b333\textunderscore515\textunderscore84051, which is M13 \#29).
On the other hand, $d_{PL}$ are on average the same 
as those obtained by M13.
% However, for b333\textunderscore215\textunderscore69147 (M13 \#19) and 
% b333\textunderscore416\textunderscore9672 (M13 \#34), our distances 
% are smaller than the literature by $\sim$1.0 and $\sim$0.7 kpc, 
% respectively; for b333\textunderscore515\textunderscore84051 (M13 \#29)
% we get the same distance, within $\sim$0.1 kpc. 
% To switch between our 
% calibration of the PL and that of M13, leads to negligible differences 
% in distance ($\lesssim$0.9 kpc) which do not affect qualitatively 
% the distance distribution.

To sum up, the effect on the distance 
estimates of the offsets in $A_{Ks}$ is 
on average that of shifting targets to higher distances. 
However, at least for T2Cs, using the PL$HK_s$ method
provides results that are similar to those of M13. Since most of these targets are 
saturated, their proper motions have large uncertainties, 
and therefore we could not decipher which 
set of $A_{Ks}$-distances is correct based on their proper motions 
(e.g., the three CCs should have disk kinematics if our distance
estimates are correct).

\begin{figure*}[!htbp]
\centering
\includegraphics[trim={0 1.3cm 0 0},clip,width=10cm]{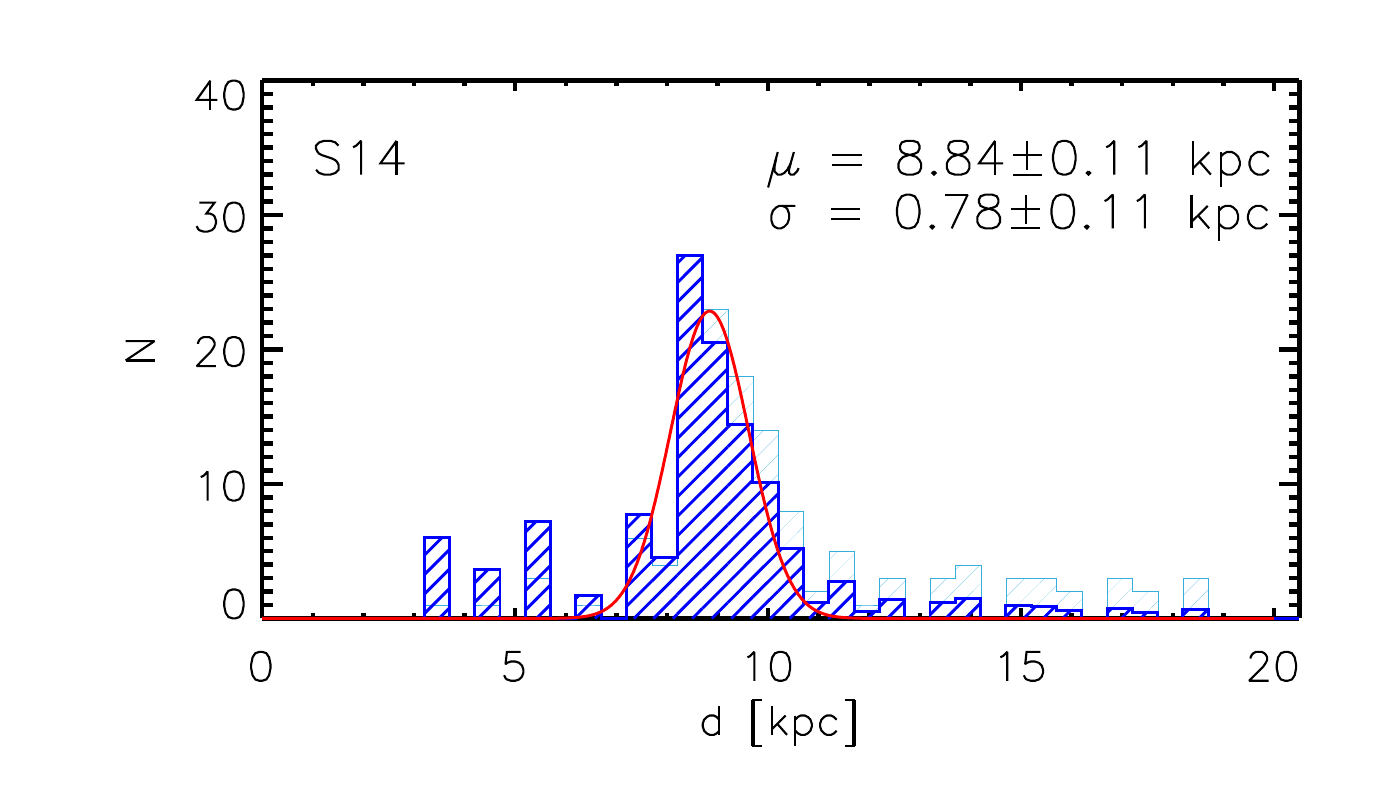}
\includegraphics[trim={0 0 0 .65cm},clip,width=10cm]{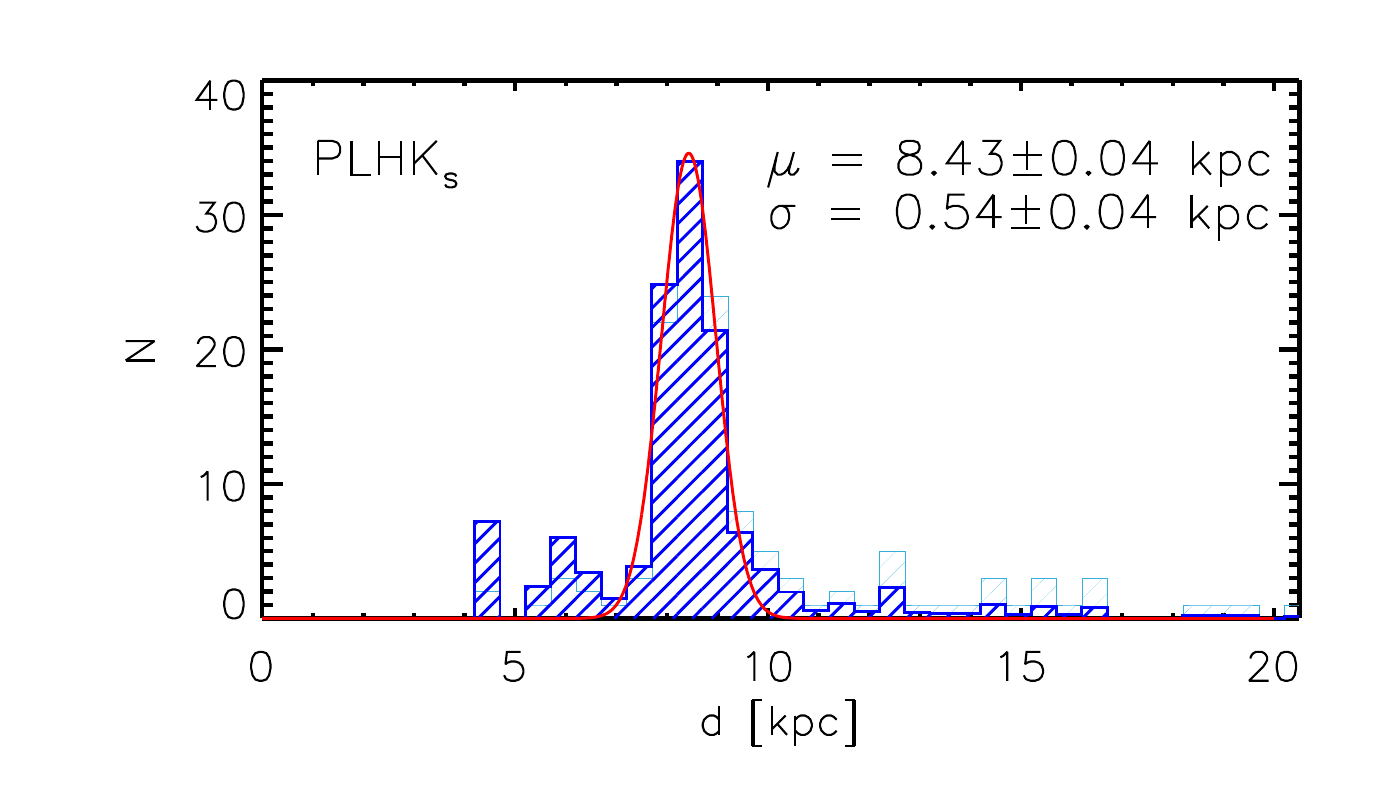}
\caption{Top: Light blue dashed histogram showing the $d_{S14}$ distribution of BLHs and WVs within 20 kpc. 
The dark blue histogram shows  the rescaled distance distribution (see text). The red solid line shows a Gaussian 
fit to the rescaled distribution. The abscissa of the peak and the 
sigma of the Gaussian are labeled. Bottom: As in top, but for $d_{PL}$.}
\label{fig:distance2}
\end{figure*}

Therefore, we inspect the 
$d_{S14}$ and $d_{PL}$ distribution of T2Cs in greater detail; 
they are displayed as light blue histograms in Fig.~\ref{fig:histoper}. 
The plain distributions provide biased information. In fact, one must take into
account the depth effect, which makes farther stars more likely to be detected.
In fact, the number of stars detected increases quadratically with distance, because
larger volumes are surveyed at larger distances. 
To take this geometric effect into account, we scaled the 
distributions by $d^{-2}$. We fitted the rescaled distributions (blue histograms in 
Figure~\ref{fig:distance2}) with Gaussians.

The peaks of the Gaussians of $d_{S14}$ and $d_{PL}$ are 8.84$\pm$0.11 kpc and 8.43$\pm$0.04 kpc, 
respectively. 
This technique is usually adopted to estimate the distance of the GC 
\citep{groenewegen08,dekany2013,pietrukowicz2015,bhardwaj17c,contreras2018,braga2018b}
but our result using the S14 map is relatively large compared to the recommended value of 
8.3$\pm$0.2(stat.)$\pm$0.4(syst.) kpc \citep{degrijs2016} and compared to the 
new geometric estimate of the distance of Sgr A* with GRAVITY 
\citep[8.127 kpc;][]{gravity2018}. Moreover, the 
bin enclosing distances between 7.5 and 8.0 kpc 
is undersampled in $d_{S14}$, making the distribution fairly asymmetrical. This is 
further evidence that the S14 map does not provide extinctions that are 
systematically offset by a fixed amount, but rather shows that there are individual 
offsets in $A_{Ks}$.
On the other hand, the distances 
obtained using the PL$HK_s$ method
are more reliable, because the peak of Gaussian is closer to the 
estimates in the literature. Finally, the 
smaller HWHM of the $d_{PL}$ distribution 
goes in the correct direction. In
fact, although investigations of the spatial distrubution 
of RRLs and T2Cs towards the entire Bulge 
generally provide larger HWHM 
\citep[$\sim$0.8-1.0 kpc][]{pietrukowicz2015,bhardwaj17c}, 
our investigation is more limited in sky area, and therefore 
we expect a tighter distribution of T2Cs.

Finally, we discuss the huge effect played by the reddening law. Not only, as
already mentioned, is the total-to-selective
absorption ratio $\dfrac{A_{Ks}}{E(J-K_s)}$ different between 
\citet{nishiyama2006,nishiyama2009} and \citet{alonsogarcia2017}, but  
a macroscopic effect is played by the extinction ratio $A_H/A_{Ks}$, 
which is 1.731 and 1.88 for the two reddening laws, respectively.
Of course the photometric systems are different, and therefore the 
comparison is not straightforward. Nonetheless, the PL$HK_s$ 
method strongly depends on this ratio and the effect is 
a systematic decrease of the distances by $\sim$1.2 kpc 
when using the \citet{nishiyama2006,nishiyama2009} 
reddening law.

All of these considerations provide strong evidence that further insight into 
the reddening law towards the central-most regions 
of the Galaxy is needed in order to correctly interpret the 
results of past and future observations, especially due to 
the uncertainty on distances and differential reddening effects.

\section{Catalog properties}\label{sect_catalog}

\begin{figure*}[!htbp]
\centering
\includegraphics[width=18cm]{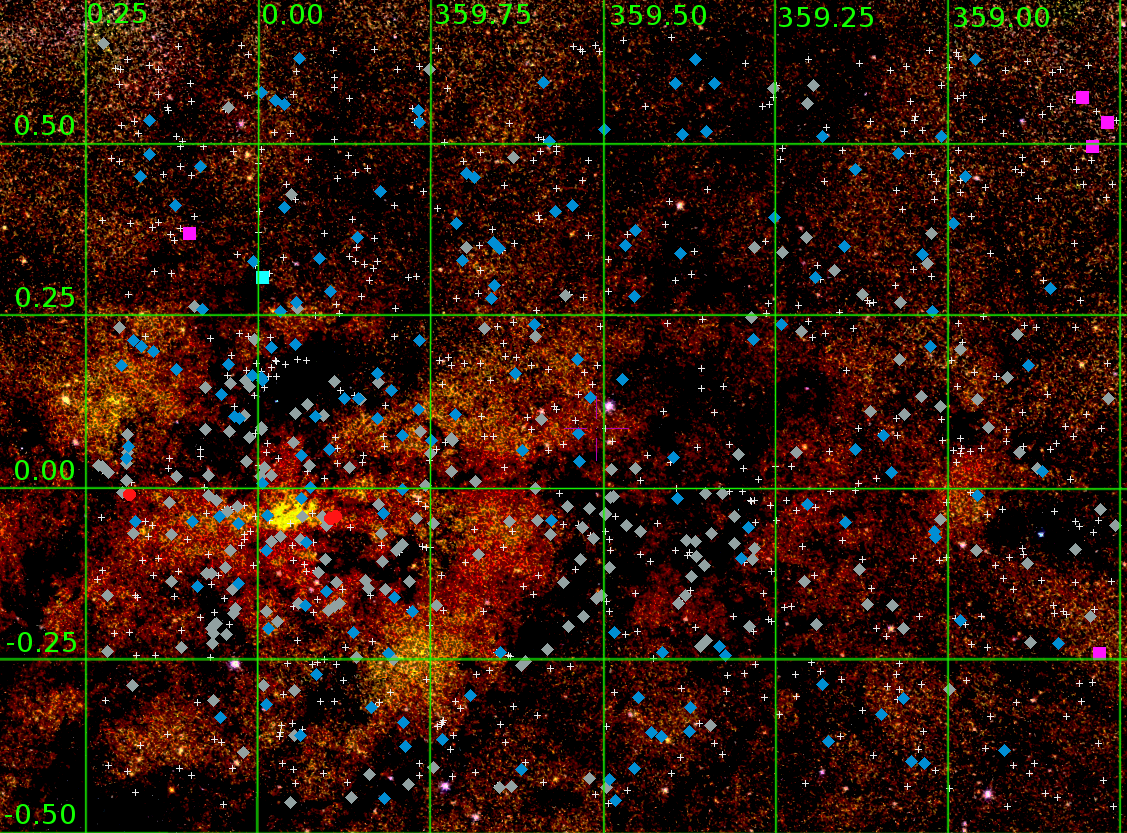}
\caption{Map in Galactic longitude and latitude ($l$,$b$) showing 
the distribution in the sky of the sample variable stars.
Map obtained with Aladin Sky Atlas v10.076 \citep{aladin}, 
using the 2MASS $JHK_s$ images catalog \citep{skrutskie2006}.
Blue diamonds represent T2Cs, which are spread over the whole field. Gray
diamonds display Miras, which are more concentrated at low latitudes.
Red ovals display CCs \citep[the ones by ][]{matsunaga11}. 
The magenta squares in this map are the RRLs, the cyan square is 
the AC, and the small white plus symbols represent all the other types of 
variables.}
\label{fig:galaxy}
\end{figure*}

The positions of the 1,019 PVSs in the Galaxy are displayed in Fig.~\ref{fig:galaxy}. 
The T2Cs (blue diamonds) are uniformly distributed within the b333 tile.
This is expected, since T2Cs are mostly old, with a 
fraction of intermediate-age stars \citep{wallerstein2002,iwanek2018}.

We also point out that the position of RRLs, which are found only 
close to the borders of the tile, is due to the intrinsic 
faintness of these objects. This is consistent with 
the spatial distribution of RRLs found by CR18, that is, they are located around 
the GC but their sky surface density decreases towards 
the GC itself (see their Figure 6).

% per fare la figura... Aladin 2MASS Galactic
% creare file di radec...
% awk '{print $64","$65","$1"_"$2}' lista_cepheids_good_temp3_v10 > radec_all_v10
%togliere header a questo
% awk '$4=="0"{print $64","$65","$1"_"$2}' lista_cepheids_good_temp3_v10 > radec_all_ccfu
% awk '$4=="2"{print $64","$65","$1"_"$2}' lista_cepheids_good_temp3_v10 > radec_all_t2c
% awk '$4=="5"{print $64","$65","$1"_"$2}' lista_cepheids_good_temp3_v10 > radec_all_rrab
% awk '$4=="3"{print $64","$65","$1"_"$2}' lista_cepheids_good_temp3_v10 > radec_all_acfu
% awk '$4=="8"{print $64","$65","$1"_"$2}' lista_cepheids_good_temp3_v10 > radec_all_mira

% attenzione, se una lista contiene un solo oggetto, metterne uno farlocco altrimenti aladin non carica

% Tool -> script console
% load /home/vittorio/Documenti/Surveys/VVV/data/tile/b333/phased_lcv/radec_all_v10
% load /home/vittorio/Documenti/Surveys/VVV/data/tile/b333/phased_lcv/radec_all_ccfu
% load /home/vittorio/Documenti/Surveys/VVV/data/tile/b333/phased_lcv/radec_all_t2c
% load /home/vittorio/Documenti/Surveys/VVV/data/tile/b333/phased_lcv/radec_all_rrab
% load /home/vittorio/Documenti/Surveys/VVV/data/tile/b333/phased_lcv/radec_all_acfu
% load /home/vittorio/Documenti/Surveys/VVV/data/tile/b333/phased_lcv/radec_all_mira

\begin{figure*}[!htbp]
\centering
\includegraphics[width=10cm]{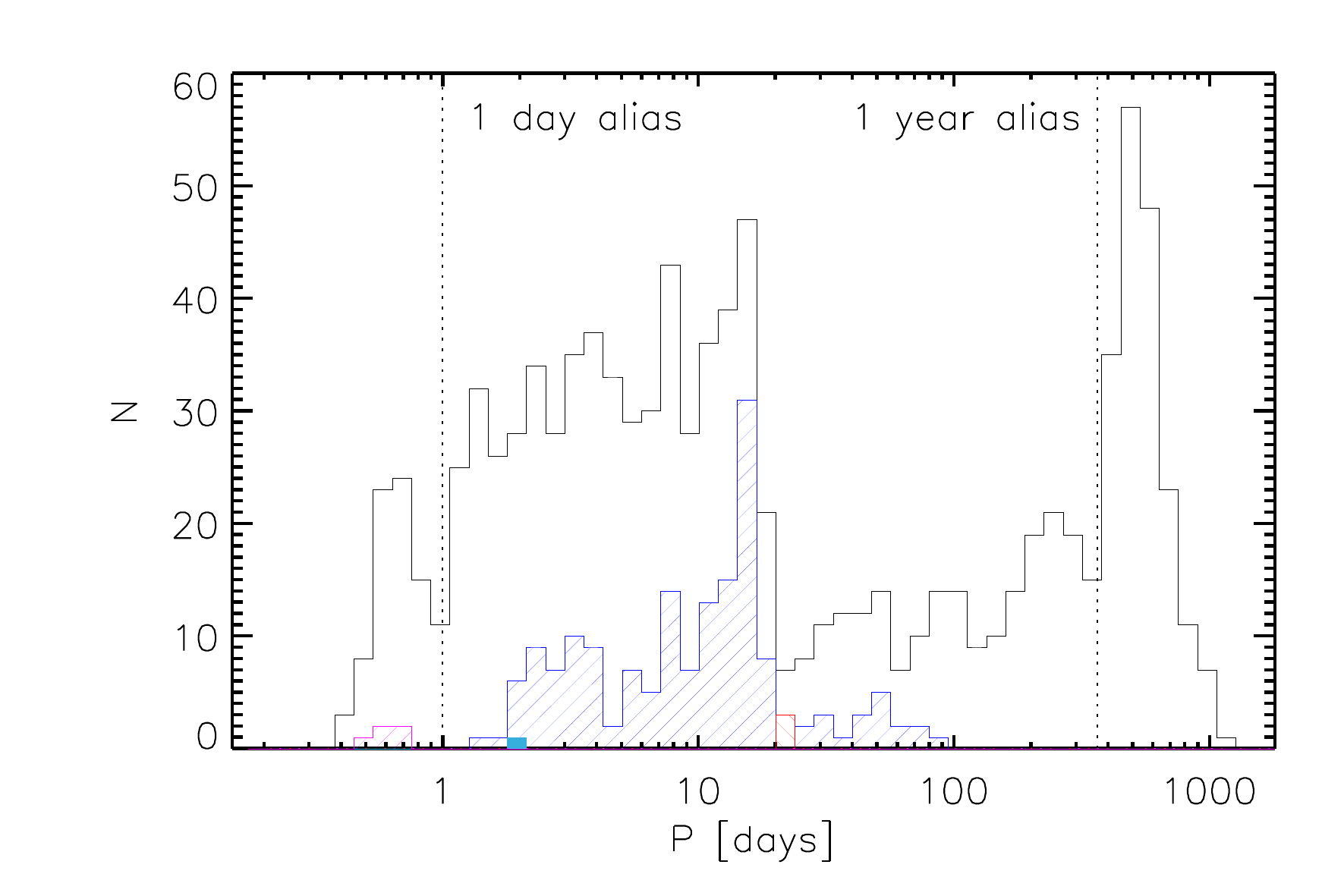}
\caption{Period distribution of our final list of PVSs. The dashed lines mark the 
aliases at 1 day and 1 year. The black histogram includes all the 1,019 candidate PVSs 
found, the dashed magenta, blue, and red histograms represent RRabs, T2Cs, and 
CCFUs, respectively. The filled light blue bin at P$\lesssim$2 days represents
the ACFU.}
\label{fig:histoper}
\end{figure*}

The distribution of the periods of our variables, displayed 
in Fig.~\ref{fig:histoper}, outlines some interesting features.
First of all, we note a lack of variables in the interval between
$\sim$20 and $\sim$365 days. The reason for this lack of stars is due 
to the aliases at one year and at half a year, and to the transition
between WVs and RVTs: the light curves of RVTs might
display alternating deep and shallow minima, 
and are therefore more difficult to classify. 

% On the other hand, we tend to believe that the lack of short-period Miras is due to the 
% intrinsic properties of the different 
% stellar populations in the surveyed area. In fact, as already mentioned,
% the periods of Miras are a proxy of their age. More quantitatively,
% for P$\gtrsim$320 days, they are younger than $\sim$7 Gyr; they approach
% an age of $\sim$3 Gyr for P$\sim$450 days and are even younger 
% for longer periods. This means that Miras trace all the 
% populations---except the youngest ($\lesssim$1 Gyr)---within the 
% surveyed area. 

Six of the nineteen RVTs display clear alternating minima. We have inspected
their position and conclude that they are not peculiar with respect to 
the other RVTs and T2Cs in general. \citep{soszynski2017} found that 
WVs with periods longer than 16 days  also might show alternating minima, but 
we did not find any of these.

\begin{figure*}[!htbp]
\centering
\includegraphics[width=12cm]{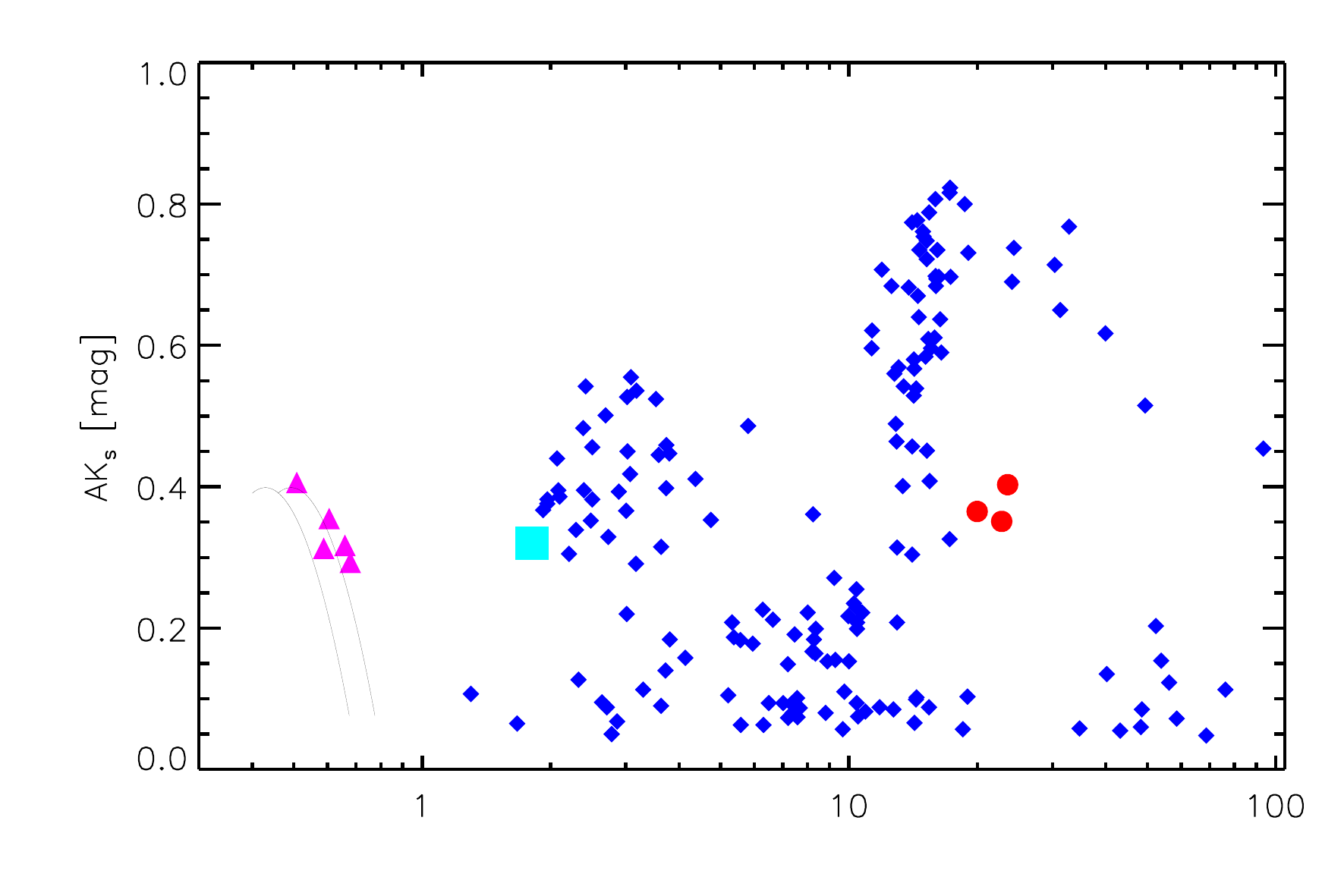}
\caption{Bailey diagram of the PVS in our final list. Small black plus symbols mark the 
position of stars with uncertain classification. Blue diamonds mark T2Cs, 
magenta triangles mark RRab stars, red circles mark CCFUs,
and the cyan square marks the ACFU. The solid lines represent
the Oosterhoff I (left) and Oosterhoff 
II (right) sequences of RRab stars \citep{cacciari05}, 
rescaled using $Amp(K_s)/AB$ amplitude ratios \citep{braga2018}.}
\label{fig:bailey}
\end{figure*}

Figure~\ref{fig:histoper} also displays a lack of variables with solid 
classification in the period between 1 and 2 days, despite the large 
number of variables (119) detected at these periods.
This is mainly due to the overlap of 
many variable types (T2Cs, CCFOs, CCFUs, ACFOs, ACFUs) 
in that region of the Bailey diagram, which is one of our 
main diagnostics for the classification. This lack of 
variables with unambiguous classification in this period range is even more 
clear in the Bailey diagram (Fig.~\ref{fig:bailey}). 

Figure~\ref{fig:bailey} also shows that the T2Cs follow the 
usual double-peaked period distribution 
\citep[][and references therein]{catelan15}. It also shows that we 
detected variables with $Amp(K_s)$ as small as 0.047$\pm$0.016 mag, 
thanks to the loose constraints described in Section~\ref{selectiontarget}.
The five RRabs, on the other hand, follow the Oosterhoff II
sequence. For the definition of the Oosterhoff groups, the 
reader is referred to \citet{oosterhoff39,catelan15}.
This is surprising, given the fact that the Oosterhoff 
classification is an indicator of 
metallicity \citep{kinman1959}, and Oosterhoff II 
RRLs should be the most metal-poor. In fact, \citet{kunder2009}
and \citet{contreras2018}, among others, 
found that bulge RRLs are much more likely to belong to the Oosterhoff 
I population. This result does not change when adopting the amplitude 
ratio by \citet{navarrete15}.

% % Test Kolmogorov-Smirnov
% import numpy as np
% import ndtest
% import matplotlib.pyplot as plt  
% 
% per1=[]
% amplk1=[]
% infile.close()
% filein="/home/vittorio/Documenti/Surveys/VVV/data/tile/b333/phased_lcv/lista_cepheids_good_temp3_allfigures_allparameters8_refined2"
% infile=open(filein,'r')
% c=0
% for line in infile:
%    if c<1:
%      c=c+1
%      continue
%    if (line.split()[2]) != '2':
%      continue
%    per1.append(float(line.split()[6]))
%    amplk1.append(float(line.split()[38]))
% 
% infile.close()
% per2=[]
% amplk2=[]
% filein="/home/vittorio/Documenti/Surveys/VVV/data/t2cep_ogle4_pvijk"
% infile=open(filein,'r')
% c=0
% for line in infile:
%    if c<1:
%      c=c+1
%      continue
%    per2.append(float(line.split()[8]))
%    amplk2.append(float(line.split()[31]))
% 
% infile.close()
% 
% amplk1_arr=np.asarray(amplk1)
% per1_arr=np.asarray(per1)
% per2_arr=np.asarray(per2)
% amplk2_arr=np.asarray(amplk2)
% 
% ind_good=np.where((amplk2_arr > -1.) & (per2_arr > 2.) & (per2_arr < 20.))
% amplk3_arr=amplk2_arr[ind_good]
% per3_arr=per2_arr[ind_good]
% 
% ind_good=np.where((amplk1_arr > -1.) & (per1_arr > 2.) & (per1_arr < 20.))
% amplk4_arr=amplk1_arr[ind_good]
% per4_arr=per1_arr[ind_good]
% 
% ndtest.ks2d2s(per3_arr,amplk3_arr,per4_arr,amplk4_arr)
% 
% fig4=plt.figure()
% plt.scatter(per3_arr,amplk3_arr)
% plt.scatter(per4_arr,amplk4_arr,c='r')
% plt.show()

\begin{figure*}[!htbp]
\centering
\includegraphics[trim={0 0 .8cm 0},clip,width=10cm]{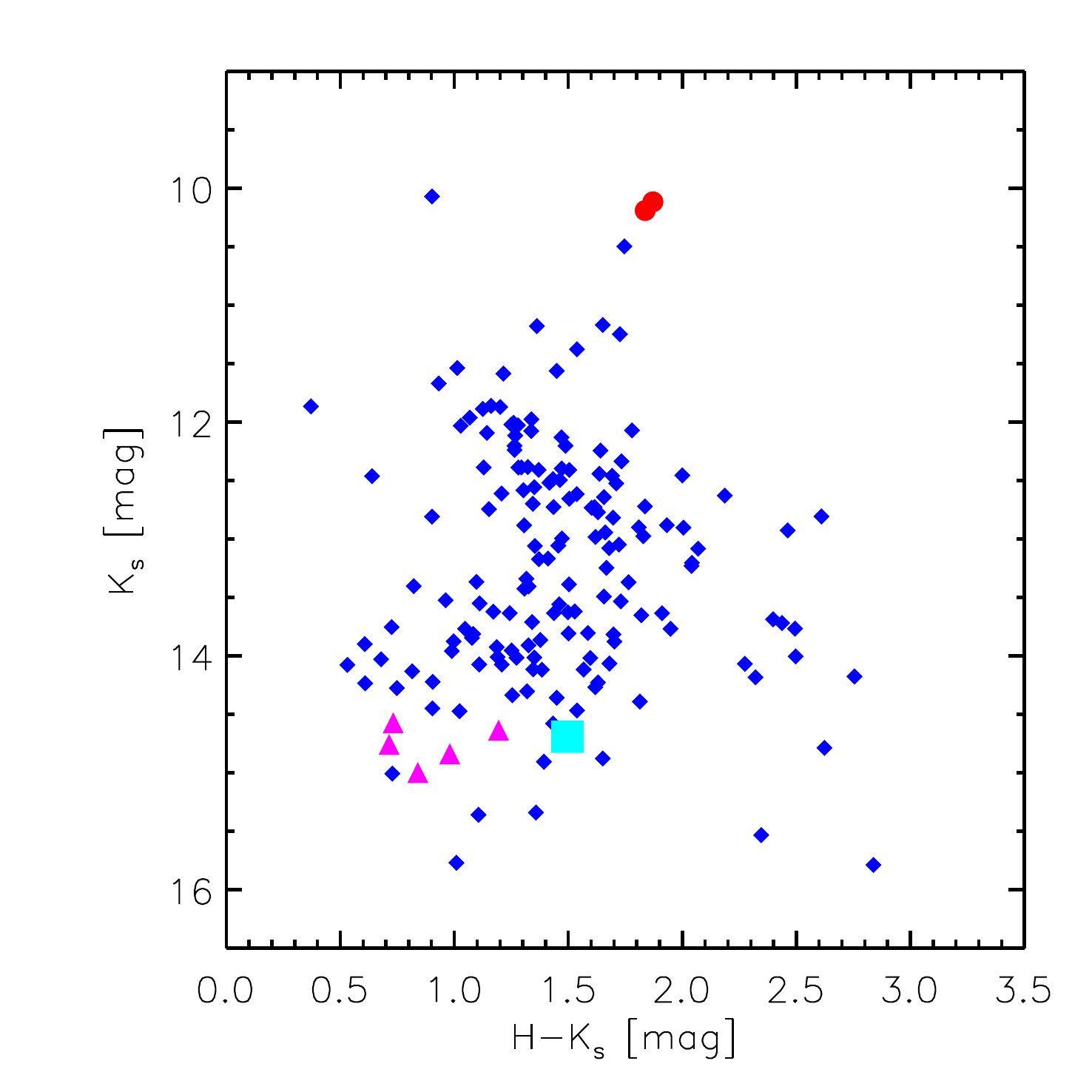}
\caption{$\langle K_s \rangle$ vs. $H-\langle K_s \rangle$ 
CMD of the candidate variable stars in our 
sample. The symbols are the same as in Fig.~\ref{fig:bailey}. The red arrow displays
the reddening vector in arbitrary units.}
\label{fig:cmd}
\end{figure*}

The observed CMD (left panel in Fig.~\ref{fig:cmd}) clearly shows the 
effects of large differential reddening in the field. In particular, 
a substantial fraction of the stars is displaced along the reddening 
vector (red arrow). 

% While some of the T2Cs may fall beyond the 
% limiting magnitude in the $K_s$-band .
% the sample of  
% Miras/LPVs should be fairly complete because their brightnesses 
% are well above the detection limit. At the bright end, however, 
% the Cepheid sample should be fairly complete, but the sample of 
% Miras/LPVs is close to saturation so we expect to miss the brightest 
% saturated stars, that should be located in the foreground.

% The reddening-corrected CMD (right panel) shows a tighter, 
% almost vertical, color distribution 
% for the T2Cs, as expected according to theri position in the instability strip. 
% The intrinsic color dispersion of the Miras/LPVs remains still large 
% after the reddening corrections, indicating some intrinsic color range for 
% these variable stars, which is also expected, due to 
% circumstellar reddening of those with periods longer than 
% $\sim$350 days, which is not taken into account . 
% The magnitude distributions of these samples of T2Cs 
% and Miras/LPVs are in agreement, within the uncertainties, with 
% the mean distance of the Bulge, 
% with a large scatter along the line of sight.

\begin{figure*}[!htbp]
\centering
\includegraphics[width=18cm]{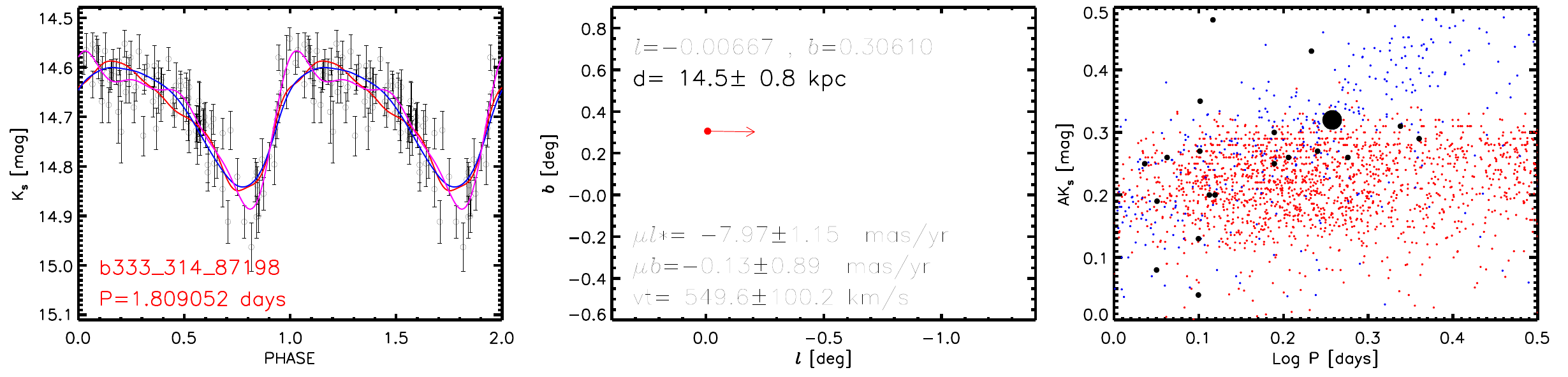}
\caption{Left: $K_s$-band Light curve of the ACFU star 
b333\textunderscore314\textunderscore87198. The name and period are labeled. The magenta line is the fourth-order Fourier fit. Blue and red lines are the template fits of T2Cs and CCs, respectively, in this
period range. Middle: Position and proper motion of the target. 
Galactic coordinates, distance, proper motion in galactic coordinates, and tangential
velocity are labeled. We note that the arrow and the tangential velocity
were derived by summing the proper motion of Sgr A* to 
our own proper motions.
Right: The large black circle represents the target in the Bailey (amplitude vs. period)
diagram. The position of both Milky Way, LMC, and SMC ACs, T2Cs and CCs are marked in black, blue, and
red, respectively.}
\label{fig:acep}
\end{figure*}

One unexpected result is the detection of the first AC in this sky area
(b333\textunderscore314\textunderscore87198, see Fig.~\ref{fig:acep}). This 
star has a period of 1.80905 days, which, as already mentioned, is within 
the range of many types of variables. However, its position in
the Bailey diagram (see right panel of Fig.~\ref{fig:acep}) restricts the 
possible classification to ACFU, T2C, and CCFU only, the latter
being very unlikely (see discussion on its tangential velocity below). However,
it is the peculiar shape of its low-noise light curve---with an extremely steep rising 
branch---which leaves no doubt on its classification. At these periods,
T2Cs and CCFU stars have more shallow light curves \citep{inno15,bhardwaj17b}. 
According to our estimates, its distance is 14.4$\pm$0.7 kpc. Also taking into account systematic
uncertainties due to reddening, there is no way that this ACFU can be within 
the Bulge. We derived a height above the Galactic plane of only 83$\pm$4 pc. 
Moreover, its proper motion has an almost null latitudinal 
component (Fig.~\ref{fig:acep}, middle panel). However, its 
tangential velocity is too small (109.2$\pm$99.4 km/s) 
to be consistent with thin disk dynamics, and therefore
it must be either a halo or a thick disk star. Spectroscopic follow-ups
of the target are planned to constrain its chemical and dynamical properties.
We have already submitted a proposal for observations with FIRE@Magellan. 
These will be useful to better constrain the complicated evolutionary 
picture of ACs, which is, to date, uncertain \citep{fiorentino12c,iwanek2018}.

\section{Conclusions}

We present the most extensive catalog of variable stars 
in the region of the GC, which includes 1,019 objects. 
The catalog contains accurate coordinates, NIR  
($ZYJH$) magnitudes, $K_s$-band mean magnitudes, 
$K_s$-band light amplitudes, and periods.
We also provide proper motions for 530 targets and
extinctions $A_{Ks}$ plus individual distances for 220 targets.
For 472 variables, we provide a high-rank, unambiguous 
classification. The latter sample includes 5 RRab, 164 T2Cs, 
3 CCFU, 1 ACFU, 16 SRVs, 210 Miras, and 73 NPVs (of which 47 DEBs).

%variabili con una distanza
% awk -F'&' '$18 !~ /ldots/ {print $18}' table2_new | wc -l

For all these candidate variables, we show the NIR CMDs, 
proper motions, period distributions,
and spatial distributions for the different types of 
variable stars, including important distance 
indicators (RRLs, Cepheids). 

{\it T2Cs} --- 164 bona fide T2Cs, among which 45 BLHs, 
100 WVs, and 19 RVTs were retrieved. Type II Cepheids are uniformly distributed
across the b333 field of view. It was not trivial to provide a 
solid classification for variables with periods between 1 and 2 days
because in that period range many possible types of variability 
overlap (T2Cs, CCFUs, CCFOs, ACFUs, ACFOs, and all types of eclipsing
binaries). We have looked in detail into their individual reddening
estimates---both individual and using extinction maps---and  
we investigated their distance distribution. We also compared
our results with those by \citet{matsunaga2013} since there are
16 T2Cs in common, but all we can conclude is that our distances 
are 1-2 kpc larger than those found in the literature. This is a very large
offset, but this should not be surprising. In fact, we estimate that 
a $\sim$10\% offset in the $A_H/A_{Ks}$ extinction ratio causes a 
shift as large as $\sim$1.2 kpc in the estimate 
for a star at the distance of the GC, 
using the same calibrating relations, $E(J-K_s)$, mean magnitudes,
and photometric system. This is a stark warning for all works 
focusing on distance estimates in regions that are close to the 
GC.

{\it Miras} --- We retrieved 210 Miras candidates with periods
between 87 and 943 days. We did not publish their astrophysical 
properties (coordinates, periods, mean magnitudes, etc.) 
because this is the aim of Nikzat et al. (2019). Nonetheless, 
we compared our photometric solution with that of 
\citet{matsunaga2009b}. The effect of the uncertainties on 
reddening are dramatic for short-period ($P<350$ days) Miras  
and turn into differences in distance as large as 10 kpc.
Invoking circumstellar extinction is not enough to justify these
differences, and both the amount of reddening and the difference in 
the adopted reddening law (\citealt{nishiyama2009} and 
\citealt{alonsogarcia2017}) play a major role in contributing to 
systematic uncertainties.

{\it CCs} --- Among the point-source catalog of our PSF photometry, we
retrieved the three CCs found by \citet{matsunaga11}. However, our 
periodicity search algorithm did not automatically detect them, since their 
light curves are relatively noisy. By adopting an a priori period, we were able to 
remove outliers from the light curves and fit their light curves to 
derive mean magnitudes and amplitudes.

{\it ACFU b333\textunderscore314\textunderscore87198} --- We found 
the first AC in this sky area. There is evidence that this 
star pulsates in the Fundamental mode and that it is located on the other side of the
Bulge. However, despite its proper motion having an almost null latitudinal 
component, its velocity is not consistent with thin-disk dynamics, and therefore
it must be either a halo or a thick disk star. 
% Spectroscopic follow-ups
% of the target are planned, and we already submitted a proposal for 
% observations with FIRE@Magellan. 

% We have started an analogous analysis of b334, which is at 
% the same latitude of b333 and is contiguous in longitude. 
% We expect to detect even more variables in b334 since the starting list
% of point sources from our photometry is 20\% larger than that of b333,
% probably thanks to a lower reddening within that region.

The published catalog is a starting point for any detailed 
investigation of variable stars within
the inner Galactic Bulge. As a matter of fact, to exploit the 
potentialities of variable stars as extinction and distance 
indicators, only a multi-band (NIR and, eventually, mid-infrared) and 
combined photometric and spectroscopic investigation 
will allow for the systematic uncertainties to be lowered. For this reason, we are
planning to submit NIR spectroscopic follow-ups of these targets
to better constrain their physical properties. 
This catalog also provides targets for future NIR spectroscopic
surveys (SDSS-V, 4MOST) and deep optical photometric surveys 
like the Large Synoptic Survey Telescope \citep[LSST,][]{ivezic2009}, 
and NIR photometric surveys like the Wide Field 
Infrared Space Telescope \citep[WFIRST,][]{wfirst1}.

% Also you could use the new distance results of Miras/LPVs from:
% Mowlavi et al. 2018  https://arxiv.org/abs/1805.02035

Acknowledgments:
We gratefully acknowledge data from the ESO Public Survey program 
ID 179.B-2002 taken with the VISTA telescope, and products from the 
Cambridge Astronomical Survey Unit (CASU). 
Support is provided by the 
Ministry for the Economy, Development and Tourism, Programa Iniciativa 
Cientifica Milenio grant IC120009, awarded to the Millennium Institute of 
Astrophysics (MAS), and by the BASAL Center for Astrophysics and Associated 
Technologies (CATA) through grant AFB-170002. 
D.M. acknowledges support from FONDECYT Regular grant No. 1170121.
M.C. and F.N, acknowledge additional support from FONDECYT regular grant
No. 1171273 and CONYCYT's PCI grant No. DPI20140066.
This research has made use of ``Aladin sky atlas'' developed at CDS, 
Strasbourg Observatory, France 

\bibliographystyle{aa}
% \bibliography{../../../Latex/ms}

\end{document}